\documentclass{article}

\usepackage{url}

\usepackage{natbib}
\bibpunct{(}{)}{;}{a}{}{,}

\usepackage{amssymb}

\usepackage{graphicx}

\usepackage{fullpage}

\usepackage{tikz}
\definecolor{grey}{rgb}{0.6,0.6,0.6}
\definecolor{darkgrey}{rgb}{0.2,0.2,0.2}

\newcommand{\sq}[0]{\tikz{\draw[color=black] (0,0) rectangle (0.3,0.3);}}

\newcommand{\grsq}[0]{\tikz{\fill[color=grey] (0,0) rectangle (0.3,0.3);}}
\newcommand{\dgrsq}[0]{\tikz{\fill[color=darkgrey] (0,0) rectangle (0.3,0.3);}}

\title
      {Reliability of relational event model estimates under sampling: how to fit a relational event model to 360 million dyadic events\thanks{Manuscript accepted for publication in \textit{Network Science}. We acknowledge support from Deutsche Forschungsgemeinschaft (DFG Grant Nr.\ LE 2237/2-1) and Swiss National Science Foundation (NSF Project Nr.\ 100018\_150126).}}

 \author        {J\"urgen Lerner\\
          University of Konstanz\\ \texttt{juergen.lerner@uni-konstanz.de}\\
         \and Alessandro Lomi\\
         University of Lugano, Switzerland\\ \texttt{alessandro.lomi@usi.ch}}

 \date{}

\begin{document}

\maketitle

\begin{abstract}
  We assess the reliability of relational event model parameters estimated under two sampling schemes: (1) uniform sampling from the observed events and (2) case-control sampling which samples non-events, or null dyads (``controls''), from a suitably defined risk set. We experimentally determine the variability of estimated parameters as a function of the number of sampled events and controls per event, respectively. Results suggest that relational event models can be reliably fitted to networks with more than 12~million nodes connected by more than 360~million dyadic events by analyzing a sample of some tens of thousands of events and a small number of controls per event. Using data that we collected on the Wikipedia editing network we illustrate how network effects commonly included in empirical studies based on relational event models need widely different sample sizes to be reliably estimated. For our analysis we use an open-source software which implements the two sampling schemes, allowing analysts to fit and analyze relational event models to the same or other data that may be collected in different empirical settings, varying sample parameters or model specification.\\

\end{abstract}


\section{Introduction}
\label{sec:intro}

Relational event networks arise naturally from directed social interaction. Examples include communication networks \citep{monge2003theories}, teams \citep{leenders2016once}, online peer-production \citep{lerner2001open,ll-tm-17}, international relations \citep{lbsb-mftien-13}, and discourse networks \citep{l-dna-16,brandenberger2019predicting}. In all these cases, social interaction reflects, and at the same time shapes, relations among social actors and affects individual sentiments -- such as trust, esteem, like, or dislike -- and their contextual behavioral expressions \citep{stadtfeld2017interactions}.

Relational events resulting from social interaction in open populations are often more directly and accurately accessible when interaction is mediated by technology, rather than embedded in face-to-face social relations. Data on relational event networks are increasingly accessible through social media platforms such as Twitter \citep{dodds2011temporal} or Facebook \citep{golder2007rhythms}. Advances in automatic data collection and coding strategies recording interaction at fine time-granularity are also encouraging the diffusion of relational event models \citep{lazer2009computational}. Relational events have been suggested as the appropriate unit of analysis, for instance, in the study of team processes \citep{leenders2016once}, and interorganizational relations \citep{vu2017relational}, where aggregating interaction over more or less arbitrary time intervals continues to represent the dominant research design \citep{pallotti2011network}. 

Building on the path-breaking work of \citet{b-refsa-08}, empirical studies adopting relational event models (REM) are becoming more common \citep{vu2017relational}. Practical applicability of REM, however, has been hampered by runtime complexity \citep{foucault2014dynamic} -- a problem that is rooted in the likelihood function which normalizes event rates on dyads that experience events by the rates on all dyads that could have possibly experienced an event at the given point in time. The corresponding \emph{risk set} can be huge, since its size is often quadratic in the number of nodes. For example, in Wikipedia -- the setting for the data that we examine in the empirical part of this paper -- more than 6 million users contribute to one or several of more than 5 million encyclopedic articles, giving rise to more than 360 million dyadic events. Thus, at given points in time a dyadic event could occur on more than 30 trillion ($3\times 10^{13}$) dyads. Computing explanatory variables (\emph{sufficient statistics}) for each of these dyads is unfeasible even for a single observed event -- much less so for all 360 million events. Such problems are rather typical since -- as mentioned above -- large quantities of information on relational events are often collected with automatic data-collection technologies which in turn can easily yield large, or very large networks.

Empirical studies applying REM, therefore, have often been limited to networks with a rather small number of nodes \citep{foucault2014dynamic,lbsb-mftien-13, vu2017relational}. Exceptions include studies exploiting a ``sparsity condition,'' namely that a single observed event changes the values of some statistics not for all dyads in the risk set but only for a limited subset \citep{perry2013point,hunter2011dynamic}. For instance, an observed event on a dyad $(a,b)$ changes the out-degree statistic only for all dyads with source $a$ -- implying update costs that are linear, rather than quadratic in the number of nodes. This approach, however, requires to develop specific estimation algorithms for different model statistics. Moreover, it does not apply to models in which statistics on all dyads change from event to event, for instance, when using ``recency statistics'' \citep{lomi2014quality,vpr-remslm-15}. However, even if sparsity conditions hold and explanatory variables change only on a fraction of the risk set, the method is still not applicable to networks of the size that we consider in this paper.

A more general and less assuming approach to alleviate the runtime problems and speed-up estimation of REM on large event-networks, involves sampling from the risk set, as suggested by \citet{b-refsa-08}. To our knowledge \citet{vpr-remslm-15} were the first to propose a concrete sampling scheme and to apply it in an empirical study (earlier mentioned in the dissertation of the first author). \citet{vpr-remslm-15} propose and apply \emph{case-control sampling} \citep{bgl-mascdcphm-95} in which a sampled likelihood is constructed by including all observed events (\emph{cases}) but only a limited number of dyads from the risk set not experiencing an event at given points in time (``null dyads'' or \emph{controls}).

\citet{vpr-remslm-15}, however, do not clarify the variation in the parameters that is caused by sampling, nor do they address issues about the number of controls that should be sampled per observed event. Similarly wanting is a discussion about the possibility of applying other sampling strategies in addition to, or instead of, case-control sampling. For instance, \citet{vpr-remslm-15} propose to sample 10,000 controls per event. On the one hand, we might wonder whether 10,000 controls is enough to obtain reliable estimates or, on the other hand, if so many controls per event are really needed. Moreover, \citet{vpr-remslm-15} analyze event networks comprising some tens of thousands of events and a maximal risk set size of 1.8 million. Since our data has more than thousand times as many events and a maximal risk set size that is more than ten million times larger, we need lower numbers of controls per event and an additional sampling strategy, besides case-control sampling, to achieve a feasible runtime.

Against the backdrop of this general discussion, this paper contributes to the practice of estimating REM by performing systematic experiments to assess the variation in the estimated model parameters caused by sampling. We consider combinations of two sampling schemes, (1) uniform sampling from the set of observed events and (2) case-control sampling, where we sample non-events from the risk set.

An alternative to these two sampling schemes would be to restrict the analysis to sub-networks that are sufficiently small, for instance, by considering only events among a specific subset of nodes. An example of this strategy is given by \citet{ll-ltar-18} who analyze edit and discussion events in Wikipedia restricted to articles about ``migration-related topics,'' which results in some $4,000$ articles and $87,000$ users connected by $950,000$ events. \citet{ll-ltar-18} applied case-control sampling from the risk set but did not sample from the observed events among this restricted set of nodes. We claim that the sampling strategy proposed and analyzed in this paper, which samples uniformly from all events, is preferable to the analysis of a sub-network for at least two reasons. First, it is questionable whether a sub-network is representative for a larger network. In the given example it seems rather plausible that articles about migration-related topics are not representative for all Wikipedia articles. Second, if we restrict the analysis to sub-networks, the computation of explanatory variables is potentially incorrect. In the example from \citet{ll-ltar-18}, Wikipedia users might contribute to articles that are not in the chosen subset so that, for instance, the out-degree, in-degree, or four-cycle statistics, see Eqs.~(\ref{eq:stat_pop}) to~(\ref{eq:stat_4cy}), might be computed incorrectly. In brief, restricting the analysis to sub-networks needs a strong assumption about the modularity or quasi-decomposability of the relational system of interest \citep{newman2006modularity,simon1996architecture,simon1977aggregation}.

In this paper we follow a different analytical strategy: even though we compute explanatory variables only for a random subset of events, \emph{all} events are aggregated into the \emph{network of past events} (see Sect.~\ref{sec:wiki_model}) which ensures that all computed explanatory variables are correct.

We contribute to the practical applicability of REM in empirical network research by:
\begin{enumerate}
\item Providing a proof-of-concept that relational event models can be fitted to large event networks with millions of nodes and hundreds of millions of events by analyzing a sample of some tens of thousands of events and a small number of controls per event. For some network effects it is even sufficient to analyze a sample comprising a few hundreds of events.
\item Assessing the variation of estimated parameters as a function of the number of sampled events and the number of controls per event, separately for different commonly-used network effects. This allows us to determine how many events and controls are needed for reliable estimation of the different effects and sheds light on the question whether we should sample few events and many controls per event or rather many events and fewer controls
\item Identifying characteristics in the data that explain why some network effects need many more observations than others to be reliably estimated and discussing a different sampling strategy that might be applied in situations of badly distributed explanatory variables.
\item Introducing an open-source software with which similar parameter estimation and sensitivity tests can be performed using different sample parameters, different models, or different data.
\end{enumerate}

The remainder of this paper is structured as follows. Section~\ref{sec:background} recalls the basic framework for relational event models and introduces the two sampling strategies. In Sect.~\ref{sec:experiments} we introduce the setting, data, empirical model specification, and the experimental design. We provide detailed results in Sect.~\ref{sec:results} and summarize their implications in Sect.~\ref{sec:discussion} where we also offer recommendations for the practice of estimating REMs and refer to an open-source software that may be adopted to replicate the results of the study in the same or different settings. We conclude with a general discussion of the implications of our work for the practice of estimating relational event models in empirical studies. The majority of the numerical results are included in an appendix.

\section{Background}
\label{sec:background}

\subsection{Relational event models}
\label{sec:rem}

A general statistical modeling framework capable of exploiting the fine-grained time information inherent in relational event sequences has been proposed by \citet{b-refsa-08}. Relational event models (REM) specify time-varying event rates for all dyads as a function of past events on the same or other surrounding dyads. Variations of this model framework include REM for weighted events \citep{bls-ness-09,ll-tm-17},  models based on marked point-processes \citep{amati2018some}, and actor-oriented REMs \citep{stadtfeld2017interactions}. 

In general, relational event models \citep{b-refsa-08} specify a likelihood for sequences of relational events $E=(e_1,\dots,e_N)$, where each event $e_i$ has the form
\[
e_i=(u_i,v_i,t_i)\enspace.
\]
In the notation above $t_i$ is the \emph{time} of the event, $u_i\in U_{t_i}$ is the \emph{source}, sender, or initiator of the event, taken from a set $U_{t_i}$ of possible source nodes at the event time, and $v_i\in V_{t_i}$ is the \emph{target}, receiver, or addressee of the event, taken from a set $V_{t_i}$ of possible target nodes at the event time. For each time point $t$ the \emph{risk set} $R_{t}\subseteq U_{t}\times V_{t}$ is the set of dyads on which an event could potentially happen at $t$. In particular, it holds that $(u_i,v_i)\in R_{t_i}$. (Note that \citet{b-refsa-08} denoted by ``support set'' what we call ``risk set''; we adopt the latter term, using the symbol $R$, since it is commonly applied in survival analysis.)

For a time point $t$, the \emph{network of past events} $G[E;t]$ is a function of events that happened strictly before $t$ (and potentially of exogenous covariates). The network of past events shapes the distribution of events at $t$ as we explain in greater detail and in the specific empirical setting of this paper in Sect.~\ref{sec:wiki_model}.

For a time point $t$ and a dyad $(u,v)\in R_t$ let $T$ denote the random variable of the time of the next event on $(u,v)$, that is the first event $e=(u',v',t')$ with $u'=u$, $v'=v$ and $t'\geq t$. The \emph{hazard rate} (also \emph{intensity} or \emph{event rate}) on $(u,v)$ at $t$ is defined by
\[
\lambda(u,v,t)=\lim_{\Delta t\to 0}\frac{Prob(t\leq T \leq t+\Delta t\,|\;t\leq T)}{\Delta t}
\enspace.
\]
Under rather weak assumptions, the hazard rate $\lambda(u,v,t)$ can be interpreted as the expected number of events in the dyad $(u,v)$ in a time interval of length one \citep{l-smmld-03}. In empirical studies one is typically interested in factors that tend to increase or decrease the dyadic event rate.

In the Cox proportional hazard model \citep{cox1972regression} -- corresponding to the REM with ordinal times in \citet{b-refsa-08} -- the hazard is specified, conditional on the network of past events $G[E;t]$ and dependent on parameters $\theta\in\mathbb{R}^k$ as
\begin{eqnarray}
  \label{eq:lambda}
  \lambda(u,v,t,G[E;t];\theta)&=&\lambda_0(t)\cdot \lambda_1(u,v,t,G[E;t];\theta)
  \enspace\mbox{, where}\\
  \label{eq:lambda_one}
  \lambda_1(u,v,t,G[E;t];\theta)&=&
  \exp\left(\sum_{\ell=1}^k\theta_{\ell}\cdot s_{\ell}(u,v,G[E;t])\right)\enspace.
\end{eqnarray}
In the equation above, $\lambda_0$ is a time-varying baseline hazard function for all dyads in the risk set and the \emph{relative event rate} $\lambda_1(u,v,t,G[E;t];\theta)$ -- a function of parameters $\theta_{\ell}$ and sufficient statistics $s_{\ell}(u,v,G[E;t])$ -- is proportional to the probability that an event at $t$ happens on $(u,v)$, rather than on any other dyad in the risk set.

In the Cox proportional hazard model the partial likelihood based on the observed event sequence $E$ is
\begin{equation}
  \label{eq:likelihood}
  L(\theta)=\prod_{e_i\in E}\frac{\lambda_1(u_i,v_i,t_i,G[E;t_i];\theta)}
  {\sum_{uv\in R_{t_i}}\lambda_1(u,v,t_i,G[E;t_i];\theta)}\enspace,
\end{equation}
and parameters $\theta$ are estimated to maximize $L$.


\subsection{Case-control sampling}
\label{sec:ccs}

The problem in maximizing the likelihood is the runtime for evaluating the denominator in each of the factors of Eq.~(\ref{eq:likelihood}). In our empirical case, the risk set at the end of the observation period has more then 30 trillion elements. Thus, we cannot even evaluate the term for the single last event -- much less the terms for all 360 million events.

A possible solution has been suggested by \citet{vpr-remslm-15} who applied case-control sampling \citep{bgl-mascdcphm-95}. In case-control sampling one analyzes all ``cases'' (dyads experiencing an event at the given point in time) but only a small random subset of the ``controls'' (dyads in the risk set not experiencing an event at the given point in time). (In this paper we will also speak of ``events'' instead of ``cases'' and of ``non-events'' instead of ``controls''.) Let $e_i=(u_i,v_i,t_i)$ be the $i$'th event in E and let $m$ be a positive integer, giving the number of controls per event. In the sampled likelihood the term $R_{t_i}$ in the denominator of Eq.~(\ref{eq:likelihood}) is replaced by $SR_i(m)\subseteq R_{t_i}$ which is drawn uniformly at random from 
\[
\{s\subseteq R_{t_i}\,\colon\; (u_i,v_i)\in s \,\wedge\, |s|=m+1\}\enspace.
\]
From another point of view,
the set $SR_i(m)$ contains the dyad $(u_i,v_i)$, that is the one in which event $e_i$ happens, and additional $m$ elements (the ``controls''), different from $(u_i,v_i)$, randomly drawn from $R_{t_i}$ without replacement. For a given sequence of sampled case-control sets $SR(m)=(SR_1(m),\dots,SR_N(m))$, the sampled likelihood is defined by
\begin{equation}
  \label{eq:likelihood_ccs}
  L^{[SR(m)]}(\theta)=\prod_{e_i\in E}\frac{\lambda_1(u_i,v_i,t_i,G[E;t_i];\theta)}
  {\sum_{uv\in SR_i(m)}\lambda_1(u,v,t_i,G[E;t_i];\theta)}\enspace.
\end{equation}

The procedure to sample from the risk set described above differs in several aspects from Markov chain Monte Carlo (MCMC) sampling as it is typically employed in the estimation of exponential random graph models (ERGM) \citep{hunter2008ergm}. One of the most crucial differences is that the risk set of a relational event model grows linearly in the number of dyads (that is, quadratic in the number of nodes), while the space of an ERGM grows exponentially in the number of dyads.

\subsection{Sampling observed events}
\label{sec:sampling_observed}
\begin{figure}
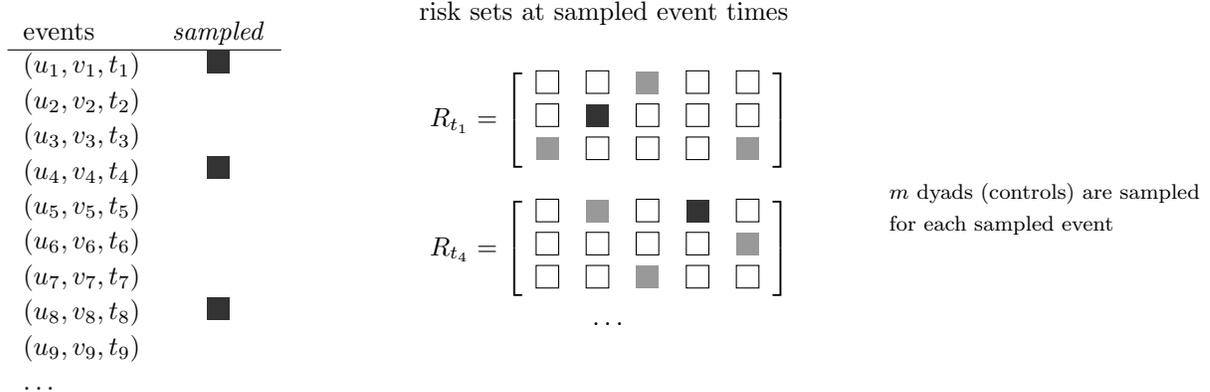

  \begin{minipage}{0.3\textwidth}
    \begin{center}
      \begin{tabular}{lc}
        events & \emph{sampled}\\
        \hline
        $(u_1,v_1,t_1)$ & \dgrsq\\[0.3ex]
        $(u_2,v_2,t_2)$ & \\[0.3ex]
        $(u_3,v_3,t_3)$ & \\[0.3ex]
        $(u_4,v_4,t_4)$ & \dgrsq\\[0.3ex]
        $(u_5,v_5,t_5)$ & \\[0.3ex]
        $(u_6,v_6,t_6)$ & \\[0.3ex]
        $(u_7,v_7,t_7)$ & \\[0.3ex]
        $(u_8,v_8,t_8)$ & \dgrsq\\[0.3ex]
        $(u_9,v_9,t_9)$ & \\[0.3ex]
        \dots & \\
      \end{tabular}
    \end{center}
  \end{minipage}\hfill
  \begin{minipage}{0.35\textwidth}
    \quad risk sets at sampled event times\medskip
    \[
    R_{t_1}=\left[\begin{array}{cccccc}
        \sq & \sq & \grsq & \sq & \sq \\
        \sq & \dgrsq & \sq & \sq & \sq \\
        \grsq & \sq & \sq & \sq & \grsq \\
      \end{array}\right] 
    \]
    \[
    R_{t_4}=\left[\begin{array}{cccccc}
        \sq & \grsq & \sq & \dgrsq & \sq \\
        \sq & \sq & \sq & \sq & \grsq \\
        \sq & \sq & \grsq & \sq & \sq \\
      \end{array}\right] 
    \]
    \begin{center}
      \dots
    \end{center}
    \vspace{0.5cm}\phantom{.}
  \end{minipage}\hfill
  \begin{minipage}{0.25\textwidth}
    {\footnotesize $m$ dyads (controls) are sampled for each sampled event}
  \end{minipage}\medskip
  
    \caption{Schematic illustration of sampling. Observed events (left-most column) are sampled independently with probability $p$; sampled observed events are indicated by black squares. For each sampled event $e_i$, we sample $m$ dyads (controls) from the risk set $R_{t_i}$; sampled controls are indicated by grey squares. Note that all observed events happening before time $t$ (not just the sampled events) are used to maintain the network of past events $G[E;t]$ and, thus, to compute the sufficient statistics $s_{\ell}$ in the specification of the event rate $\lambda_1$ in Eq.~(\ref{eq:lambda_one}).}
    \label{fig:sampling}
  \end{figure}

Another sampling strategy that we consider in this paper additionally to case-control sampling involves uniform sampling from the set of observed events. Concretely, for a real number $p$ with $0 < p \leq 1$, let $SE(p)\subseteq E$ be a random subset of observed events, obtained by including each $e\in E$ independently at random with probability $p$. Additionally, let $SR(m)$ be a sequence of sampled case-control sets as described above. The sampled likelihood is given by:
\begin{equation}
  \label{eq:likelihood_sampling_observed}
  L^{[SR(m),SE(p)]}(\theta)=\prod_{e_i\in SE(p)}\frac{\lambda_1(u_i,v_i,t_i,G[E;t_i];\theta)}
  {\sum_{uv\in SR_i(m)}\lambda_1(u,v,t_i,G[E;t_i];\theta)}\enspace,
\end{equation}
and parameters $\theta$ are estimated to maximize the sampled likelihood $L^{[SR(m),SE(p)]}(\theta)$. Both sampling schemes, case-control sampling and uniform sampling from the observed events, are illustrated in Fig.~\ref{fig:sampling}.

We emphasize that the networks of past events $G[E;t_i]$ in Eq.~(\ref{eq:likelihood_sampling_observed}) are functions of all events in $E$ that happen before $t_i$, not just functions of the sampled events in $SE(p)$. Thus, for a given dyad $(u,v)$ the explanatory variables (sufficient statistics) $s_{\ell}(u,v,G[E,t])$ from Eq.~(\ref{eq:lambda_one}) are computed correctly. This points to a crucial difference to the approach that analyzes sub-networks as discussed in the introduction. Moreover, it points to a crucial difference to situations where events are \emph{missing} (rather than left out due to sampling); obviously, missing events cannot be used to compute statistics.

Using the sampled likelihood for estimation of model parameters introduces another source of uncertainty: if we repeated the sampling of events and controls, we obtained a different sampled likelihood and thus, probably, also different parameter estimates. In our experiments, which are explained in detail in Sect.~\ref{sec:wiki_experimental_design}, we want to determine how the variation in the model parameters $\theta$ introduced by sampling depends on the sampling parameters $p$ (the probability to sample an observed event) and $m$ (the number of sampled controls per sampled event). In this paper we call $p$ and $m$ the \emph{sample parameters} and $\theta$ the \emph{model parameters} -- unless it is clear from context which set of parameters is meant.

\subsection{Asymptotic properties}
\label{sec:aysmptotic}

An attractive property of case-control sampling in the Cox proportional hazard model is that it yields an estimator which is consistent and asymptotically normal \citep{bgl-mascdcphm-95}. This is a non-trivial result since case-control sampling is a sampling strategy that actually manipulates the ratio of events to non-events. These asymptotic properties are reassuring -- but they do not diminish the relevance of the experimental approach of this paper.

In our experiments we assess the variation in the parameters caused by sampling for different sample sizes. The experimental results we present suggest that for some of the most common network effects typically included in empirical model specifications \citep{lomi2014quality}, variation caused by sampling is much larger than the estimated standard errors of model parameters, even for relatively large sample sizes. It seems difficult to determine the sufficient sample size by purely theoretical considerations based on asymptotic properties -- which is why we consider experimental evaluation as necessary.

\section{Experiments}
\label{sec:experiments}

\subsection{Empirical setting: Network analysis of collaboration in Wikipedia}
\label{sec:wiki_setting}

Wikipedia is a prime example of a self-organizing online production community in which voluntary users invest personal time and effort to provide a public good \citep{lerner2001open,ll-tm-17}. The most crucial resources that articles need to garner in order to attain a high level of quality are the attention and contribution of users \citep{lerner2018knowledge}. This observation motivates a study of the factors that induce particular users to contribute more or less to particular articles.  The network of relational events that we consider in this paper includes all events in which any registered user contributes to any article in the English-language edition of Wikipedia. The estimated model reveals basic network effects that explain dyadic contribution rates on all user-article pairs.

Previous work analyzing user contributions in Wikipedia mostly aggregates edits over users, articles, or time. For instance, \citet{moat2013quantifying} analyze patterns in the dynamics of edits received by articles -- but do not model \emph{dyadic} event rates, nor do they analyze network effects explaining edit frequencies. \citet{yasseri2012circadian} analyze periodic patterns in the global edit activity in different language editions of Wikipedia. \citet{keegan2012editors} fit exponential random graph models to the two-mode network connecting Wikipedia users to ``breaking-news articles''. This model framework, however, requires to aggregate relational events per dyad over time. 

Surprisingly -- considering that Wikipedia contribution data are naturally provided as time-stamped relational events -- studies analyzing Wikipedia collaboration networks as relational event systems are still relatively infrequent. Perhaps this is due to the lack of methods that are able to estimate REM for large event networks. \citet{ll-tm-17,lerner2018free,ll-nsscw-19} propose and estimate REMs for the edit networks of single articles. Their model does not explain user activity rates but rather specifies dyadic probabilities to undo or redo contributions of other target users, given that a source user becomes active on the focal article. Previous work that seems to be the closest to this paper is a recent article of \citet{ll-ltar-18} which analyzes a REM for edit and talk events on Wikipedia articles about migration-related topics (also discussed in the introduction). Thus, \citet{ll-ltar-18} analyze a thematically delineated sub-network. They apply case-control sampling but do not sample from the observed events and do not experimentally vary sample sizes. Data on the Wikipedia editing network provide a particularly valuable setting to perform the experiments that we propose in this paper because of their size, accuracy and complexity, and because of the substantive interest in Wikipedia as one of the largest and most successful examples of peer production projects \citep{jan2017testing,greenstein2012wikipedia}.  

\subsection{Data}
\label{sec:wiki_data}

The event network that we consider in this paper is given by the sequence of relational events in which any registered and logged-in Wikipedia user uploads a new version of any article in the English-language edition of Wikipedia in the time period from January 15th 2001 (the launch of Wikipedia) to January 1st, 2018 (the time of data collection). For a point in time $t$ we define $U_t$ to be the set of all users who edited at least one article at or before $t$ and we define $A_t$ to be the set of all articles that received at least one edit at or before $t$. The risk set at time $t$ is defined to be the full Cartesian product of users and articles $R_t=U_t\times A_t$. It is possible to define the risk set in more sophisticated ways: users or articles could enter or leave the risk set as a function of events, due to extended periods of inactivity, or due to exogenously given entry-times or exit times. Such more complex risk set dynamics, however, are not considered in this paper. (We performed robustness checks by letting users and articles drop out of the risk set due to extended periods of inactivity; results were not affected in any meaningful way.)

Wikipedia makes its complete database available for public use.\footnote{\url{https://dumps.wikimedia.org/}} All edit events can conveniently be extracted from the files \texttt{enwiki-<date>-stub-meta-history.xml.gz} which contain meta-information (including user, article, and time) of every edit but not the text of the different versions (which we do not need for the analysis in this paper).

The time of edits in the Wikipedia data is accurate to the second. The Cox proportional hazard model does not make use of the precise timing of events. The given timestamps of events are used to determine the decay of past events, as explained in Sect.~\ref{sec:wiki_model}, and the order of events, but have no further influence on the model. In the given data several events can occur in the same second of time. We nevertheless imposed a strict order on all events. If two events happen simultaneously, then the order in which they appear in the input file was taken as the given order. Two non-simultaneous events are ordered by their timestamps. We believe that the decision to enforce a strict order has no significant impact on the results (actually it is very unlikely to select two simultaneous events in the same sample of events).

The data consists of $361,769,741$ dyadic events of the form $e=(u,a,t)$ indicating that user $u\in U_t$ edits article $a\in A_t$ at time $t$. At the end of the observation period $U_t$ contains $6,701,379$ users and $A_t$ contains $5,542,465$ articles. Thus, at the end of the observation period the risk set contains $6,701,379 \cdot 5,542,465 \approx  3.714\cdot 10^{13}$ (approximately 37 trillion) user-article pairs. For reproducibility and for facilitating similar research, the preprocessed data is available at \url{https://doi.org/10.5281/zenodo.1626323}.

\subsection{Model}
\label{sec:wiki_model}

We specify the relative event rates $\lambda_1(u,v,t,G[E;t];\theta)$ on all user-article pairs $(u,a)\in R_t$ at time $t$ as a function of past events that happened strictly before $t$, see Eq.~(\ref{eq:lambda_one}). In doing so, we weight events that happened further in the past weaker than events that happened more recently.

More precisely, following the model proposed in \citet{bls-ness-09}, we define the \emph{network of past events} $G[E;t]$ at time $t$ as the weighted two-mode network with node sets $U_t$ and $A_t$ where the weight $w(u,a,t)$ on any $(u,a)\in U_t\times A_t$ is a function of past events happening in $(u,a)$ before $t$, that is, the weight is a function of the events
\[
E_{uat}=\{(u',v',t')\in E\,\colon\;u'=u\wedge a'=a \wedge t'<t\}
\]
given by the formula
\begin{equation}\label{eq:weights}
w(u,a,t)=\sum_{(u,v,t')\in E_{uat}}\exp\left(-(t-t')\cdot\frac{\ln(2)}{T_{1/2}}\right)
\end{equation}
for a given halflife $T_{1/2}>0$. 

We set $T_{1/2}$ to 30 days so that an event counts as (close to) one in the very next instant of time, it counts as $1/2$ one month later, it counts as $1/4$ two months after the event, and so on. To reduce memory consumption needed to store the network of past events, we set a dyadic weight to zero if its value drops below $0.01$. If a single event occurred in some dyad this would happen after $6.64\cdot T_{1/2}$, that is after more than half a year. The weight $w(u,a,t)$ can be interpreted as the recent event intensity in the dyad $(u,a)$ before $t$. \emph{Edges} in the network of past events are implicitly defined to be all dyads with non-zero weights. As a robustness check we repeated some of our analyses in a model without any decay (not reported in this paper); in general, findings did not change in any meaningful way.

Following \citet{ll-ltar-18}, the event rate $\lambda_1(u,v,t,G[E;t];\theta)$, see Eq.~(\ref{eq:lambda_one}), is specified using the following five sufficient statistics.
\begin{eqnarray}
  \label{eq:stat_rep}
  repetition(u,a,G[E;t]) & = & w(u,a,t)\\
  \label{eq:stat_pop}
  popularity(u,a,G[E;t]) & = & \sum_{u'\in U_t}w(u',a,t)\\
  \label{eq:stat_act}
  activity(u,a,G[E;t]) & = & \sum_{a'\in A_t}w(u,a',t)\\
  \label{eq:stat_4cy}
  four.cycle(u,a,G[E;t]) & = & \sum_{u'\in U_t\setminus\{u\}}\sum_{a'\in A_t\setminus\{a\}}\min[w(u,a',t),w(u',a',t),w(u',a,t)]\\
  \label{eq:stat_ass}
  assortativity(u,a,G[E;t]) & = & popularity(u,a,G[E;t])\cdot activity(u,a,G[E;t])
\end{eqnarray}

The statistic $repetition(u,a,G[E;t])$ characterizes the recent event intensity in the same dyad $(u,a)$. A positive parameter associated with this statistic would reveal that users are more likely to edit articles they have recently edited before. The popularity statistic aggregates the recent event intensities in all dyads with target $a$, that is, it gives the recent intensity by which the article $a$ is edited. A positive parameter associated with the popularity statistic would reveal that popular articles (those, that recently received many edits by any user) are likely to receive edits (by potentially different users) at a higher rate in the future. The activity statistic aggregates the recent event intensities in all dyads with source $u$, that is, it gives the recent cumulative edit activity of user $u$. A positive parameter associated with the activity statistic would reveal that users who have recently been more active will also edit (potentially different articles) at a higher rate in the future. The four-cycle effect models local clustering. If users $u$ and $u'$ recently collaborated in writting an article $a'$ and, in addition, user $u'$ recently contributed to article $a$, then a positive parameter associated with $four.cycle(u,a,G[E;t])$ would reveal that $u$ has a tendency to edit $a$ at a higher rate. The assortativity statistic models the interaction between user activity and article popularity. A positive parameter would reveal that the more active users are rather drawn to editing the more popular articles while a negative assortativity parameter would reveal that the more active users are rather drawn to editing the less popular articles -- where these relations have to be understood relative to the effects of the basic variables activity and popularity.

These five effects are very common in relational event modeling and are not particular to the specific setting that we examine -- except that when analyzing a one-mode network, instead of a two-mode network, we would typically include a three-cycle effect, perhaps instead of the four-cycle effect. These effects are interesting for our current study since they assume dependence of dyadic observations on previous observations in varying levels of complexity. Repetition assumes that the rate of future events in $(u,a)$ depends just on previous events in the same dyad. The popularity and activity effects assume that the rate of future events in $(u,a)$ depends on previous events that have the same target article $a$ (for popularity) or the same source user $u$ (for activity). The four-cycle effect assumes that the rate of future events in $(u,a)$ depends on previous events in a three-path of the form $(u,a'),(a',u'),(u',a)$ for any user-article pair $(u',a')$ -- that are different from $u$ and $a$, respectively -- provided that there was at least one recent event in all three dyads $(u,a')$, $(u',a')$, and $(u',a)$. Assortativity is an interaction effect modeling that activity of a user $u$ has a different impact for editing popular articles than for editing less popular articles and vice versa. Considering these five effects in our study is interesting since it is plausible that the different effects need different sample sizes to be reliably estimated.

Due to skewed distributions we apply the mapping $x\mapsto \log(1+x)$ to the four statistics $repetition$, $popularity$, $activity$, and $four.cycle$. The interaction effect $assortativity$ is defined to be the product of $popularity$ and $activity$ after applying this logarithmic transformation.

Table~\ref{tab:cov_cor} provides information on correlation, variance, and covariance of the five statistics computed on the aggregated observations from 100 samples with sample parameters $p=10^{-4}$ and $m=5$ (see Sects.~\ref{sec:wiki_experimental_design} and~\ref{sec:results_fix}). All correlations are positive and moderately high.

\begin{table}
  \caption{Correlation (above the diagonal), variance (bold, on the diagonal), and covariance (below the diagonal) of the five statistics computed from about 21 million observations obtained by aggregating 100 samples with $p=10^{-4}$ and $m=5$.}
  \label{tab:cov_cor}
\centering\begin{tabular}{l@{\quad}r@{\quad}r@{\quad}r@{\quad}r@{\quad}r}
  \hline
 & repetition & popularity & activity & four.cycle & assortativity \\ 
  \hline
repetition & \bf 0.33 & 0.64 & 0.41 & 0.35 & 0.72 \\ 
  popularity & 0.36 & \bf 0.94 & 0.42 & 0.49 & 0.75 \\ 
  activity & 0.60 & 1.02 & \bf 6.22 & 0.79 & 0.70 \\ 
  four.cycle & 0.27 & 0.64 & 2.65 & \bf 1.83 & 0.76 \\ 
  assortativity & 2.25 & 3.95 & 9.37 & 5.55 & \bf 29.17 \\ 
   \hline
\end{tabular}
\end{table}

\subsection{Runtime complexity}
\label{sec:runtime}

The total time for computing all explanatory variables is the computation time for maintaining the network of past events plus the time needed to compute the statistics for all sampled observations, that is for sampled events and sampled controls. Below we briefly discuss that maintenance of the network of past events has a much smaller asymptotic runtime per event than the computation of statistics per dyad. This is an important property of our approach since we use all events to maintain the network of past events but we compute statistics only on a sample of observations.

The network of past events can be updated in constant time per event, so that the total cost in maintaining the network of past events is proportional to the total number of events $|E|$. This might seem surprising since the weights on \emph{all} edges change due to decay whenever time increases, see Eq.~(\ref{eq:weights}) -- which actually happens after most input events. We achieve the linear runtime by a ``lazy'' strategy: we do not perform the time-decay when time increases; instead we store for each edge, additionally to its weight, the time of the last weight-update. The weight on a dyad $(u,a)$ is only updated if either an event happens in $(u,a)$ or if the value $w(u,a,t)$ has to be accessed to compute statistics for $(u,a)$ or surrounding dyads.

The cost for computing the statistics varies. The values of $repetition$, $popularity$, $activity$, and $assortativity$ for a given dyad at a given point in time can be computed in constant time. In order to achieve this for the degree statistics ($popularity$ and $activity$) we maintain weighted in-degrees of articles and weighted out-degrees of users explicitly as functions defined on nodes in the network of past events. (The cost for maintaining these node-weights is also linear in the number of events.) The time for computing the four-cycle statistic of a dyad $(u,a)$ is bounded by the product of the unweighted degrees (that is, the number of incident edges) of $u$ and $a$. If the unweighted degrees of all nodes were equal to $d$, then the runtime for computing the four-cycle statistic for one dyad would be bounded by $d^2$.

In summary, it is much more costly to compute statistics for one observation than to update the network of past events as a result of one event. This is the rationale for our approach to use all events for maintaining the network of past events but only a fraction of events to compute the likelihood defined in Eq.~(\ref{eq:likelihood_sampling_observed}). As discussed before, taking all events to update the network of past events ensures that the computed statistics for each sampled observation are correct.

\subsection{Experimental design}
\label{sec:wiki_experimental_design}

The general goal of the computational experiments we report in the next section, is to determine how the distributions of estimated model parameters $\theta$ -- and their associated standard errors, z-values, and significance levels -- depend on the sample parameters $p$ and $m$. We first conducted a preliminary exploratory analysis to identify sample parameters $p_0$ and $m_0$ that are large enough to reliably assess the effects of all five statistics. It turned out that the different statistics require largely different numbers of observations to be reliably estimated (discussed in more detail in Sect.~\ref{sec:results}). We have chosen $p_0$ and $m_0$ such that at least the direction of all five effects, that is, whether they tend to increase or decrease the event rate, can be assessed with high confidence. After the preliminary analysis we set these initial sample parameters to $p_0=0.0001=10^{-4}$ and $m_0=5$ resulting in approximately $36,000$ sampled events and about $5\cdot 36,000$ sampled controls, that is about $216,000$ sampled observations. Then we performed four types of experiments to assess how the distributions of estimated model parameters depend on the sample parameters.

\begin{itemize}
\item \textbf{(fixed $p$ and $m$)} We repeat the sampling one hundred times with fixed sample parameters $p=p_0$ and $m=m_0$. Doing so allows us to assess for each of the five effects the variation of model parameters, standard errors, and z-values caused by sampling.
\item \textbf{(varying $p$ for fixed $m$)} For fixed $m=m_0$ we vary $p$ as in $(p_0/2^i)_{i=1,\dots,10}$. That is, for $i$ from one to ten we halve, starting from $p_0$, the probability to sample events in each step. This results in approximately $18,000$ sampled events for $i=1$ and only about 35 sampled events for $i=10$. The number of sampled controls is five times the number of sampled events, so that the number of sampled observations ranges from $108,000$ to $211$. For each $i=1,\dots,10$ we repeat the sampling ten times to assess the distribution of estimated model parameters for the five effects. In this type of experiment we assess which sample sizes are sufficient to reliably estimate parameters for the various effects.
\item \textbf{(varying $m$ for fixed $p$)} For fixed $p=p_0/10=10^{-5}$ we vary $m$ as in the sequence $(2^i)_{i=0,\dots,7}$. This results in one control per event for $i=0$ and $128$ controls per event for $i=7$. The expected number of sampled events is fixed to about $3,600$ so that the number of sampled observations ranges from $7,200$ (for $i=0$) to $464,400$ (for $i=7$). Thus, for $i=7$ we sample more than twice as many observations as for the initial sample parameters $p_0$ and $m_0$ -- but the sampled observations contain much fewer events and a much higher ratio of controls per event. For each $i=0,\dots,7$ we repeat the sampling ten times. In this type of experiment we assess the benefit of using more controls per event for a given limited number of events.
\item \textbf{(varying $p$ and $m$ for a fixed budget of observations)} Similar to the experiment type above, we vary $m$ as in the sequence $(2^i)_{i=1,\dots,8}$. However, in this type of experiment we let $p$ change as a function of $m$ to fix the number of sampled observations to that determined by $p_0$ and $m_0$. More precisely, for $i=1,\dots,8$ we set $m=2^i$ and $p=6\cdot 10^{-4}/(2^i+1)$. The expected number of sampled observations is, thus, fixed at about $216,000$. The number of controls per event ranges from 2 to $256$ and the expected number of sampled events ranges from about $72,000$ (for $i=1$) to about $840$ (for $i=8$). For each $i=1,\dots,8$ we repeat the sampling ten times. In this type of experiment we assume that we have a given budget of observations (for instance, constrained by runtime considerations) and assess whether it is preferable to include more events (and fewer controls per event), or the other way round, or if there is an optimal intermediate number of controls per event.
\end{itemize}

After conducting these experiments and analyzing the results we perform additional analyses to identify reasons why the estimation of the model parameter for the repetition effect needs so many more observations than for the other effects. 

\section{Results}
\label{sec:results}

In Sect.~\ref{sec:results_single} we present and discuss estimated parameters on a single sample obtained with $p=p_0$ and $m=m_0$. Sections~\ref{sec:results_fix} to~\ref{sec:results_co_vary} present and discuss results of the four types of experiments described above. Section~\ref{sec:results_repetition} conducts additional analyses to clarify findings about the repetition effect.

\subsection{Model estimation on a single sample}
\label{sec:results_single}

Table~\ref{tab:single} reports estimated model parameters and standard errors of a single sample obtained with the initial sample parameters $p_0=10^{-4}$ and $m_0=5$ resulting in about $36,000$ events and $216,000$ observations. All parameters are significantly different from zero at the $0.1\%$ level, point in the expected direction, and are consistent with results reported in previous studies based on a much smaller sub-network \citep{ll-ltar-18}.

\begin{table}
  \caption{Estimated parameters and standard errors (in brackets) on a single sample obtained with $p_0=0.0001$ and $m_0=5$. All parameters are significantly different from zero at the 0.1\% level.}
  \label{tab:single}
  \centering\begin{minipage}{0.45\textwidth}
\begin{tabular}{l@{\quad}r }
\hline
repetition                         & $12.348 \; (0.999)$ \\
popularity                & $1.244 \; (0.035)$  \\
activity                     & $1.445 \; (0.030)$  \\
four.cycle                        & $0.659 \; (0.089)$  \\
assortativity & $-0.120 \; (0.024)$ \\
\hline
AIC                                & 5,607.475                  \\
Num. events                        & 36,114                     \\
Num. obs.                          & 216,684                    \\
\hline
\end{tabular}
  \end{minipage}
\end{table}

The parameter associated with $repetition$ is positive indicating that a user $u$ typically is more likely to edit an article if she has recently edited it before. We find a positive popularity effect: articles that recently received more edits from any user are likely to be edited at a higher rate by potentially different users. Considering that contributions by users are a crucial resource for articles \citep{lerner2018knowledge}, this ``rich-get-richer'' effect \citep{ba-esrn-99} might actually have adverse implications for the development and quality of Wikipedia. Indeed, if contributing users tend to crowd together on a few popular articles, other articles are in danger of being neglected. We also find a positive activity effect: users that recently contributed more frequently to any article are likely to edit potentially different articles at a higher rate. The positive parameter of the four-cycle effect points to local clustering, where the clusters might be interpreted as latent topics of articles, or interests of users, respectively. If the interpretation via user interests and article topics applies, the four-cycle effect can be understood as follows. If two users $u$ and $u'$ recently co-edited an article $a'$, then they are likely to have similar interests. If, additionally $u'$ recently edited another article $a$, then $a$ and $a'$ are likely to involve similar topics. In turn user $u$ is also likely to edit article $a$ at a higher rate.
Assortativity, that is the interaction effect between the activity of users and the popularity of articles, is found to be negative. Thus, while users in general have the tendency to edit already popular articles, this behavior is less pronounced for the more active users. Thus, the more active users seem to dedicate a bigger share of their work (than the less active users) to the less popular articles. In the light of the discussion about the popularity effect above, this negative assortativity might be very important for Wikipedia, since it gives an indication that the more active users might be the ones who start editing less popular articles -- which would otherwise be in danger of being neglected.

Having established the consistency of the results with prior studies \citep{ll-ltar-18}, and the contextual interpretability of the empirical estimates, we focus in the next sections on the variation in the estimates associated with different sample parameters.

\subsection{Variation caused by sampling (fixed sample parameters)}
\label{sec:results_fix}

\begin{figure}
    \includegraphics[width=0.2\textwidth]{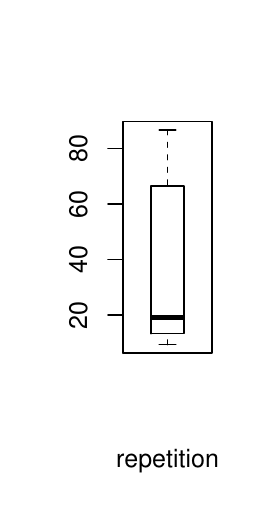}\hfill
    \includegraphics[width=0.2\textwidth]{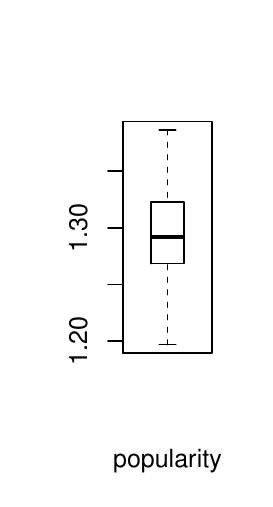}\hfill
    \includegraphics[width=0.2\textwidth]{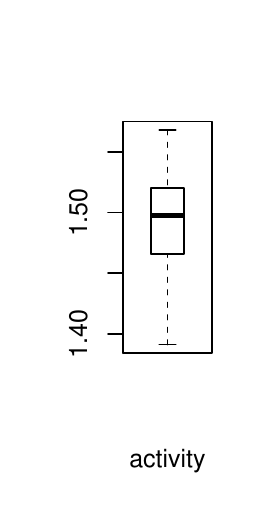}\hfill
    \includegraphics[width=0.2\textwidth]{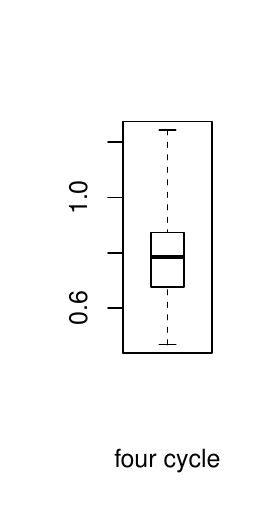}\hfill
    \includegraphics[width=0.2\textwidth]{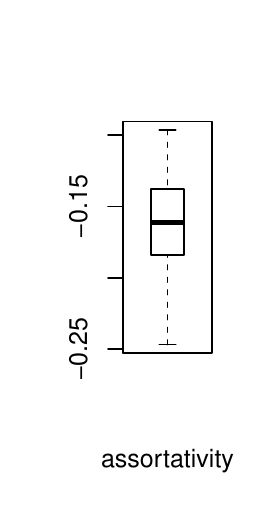}
    \caption{Boxplots (different $y$-axes) of parameter estimates for the five effects over 100 samples with $p=0.0001$ and $m=5$.}
    \label{fig:boxplots_fix}
  \end{figure}

Table~\ref{tab:summary_fix} reports summary statistics of estimated parameters, standard errors, and z-values for the five effects over these 100 samples. The boxplots in Fig.~\ref{fig:boxplots_fix} display the summary statistics for the estimated parameters. 

\begin{table}
  \caption{Summary statistics (minimum, first quartile, median, mean, third quartile, maximum, and standard deviation) of parameters (\emph{par}), standard errors (\emph{se}), and z-values (\emph{z}) for the five effects repetition (\emph{rep}), popularity (\emph{pop}), activity (\emph{act}), four.cycle (\emph{4cy}), and assortativity (\emph{asr}) over 100 samples with $p=0.0001$ and $m=5$. Bold-font highlights the standard deviation of the parameters and the mean of the standard errors.}
  \label{tab:summary_fix}
\centering\begin{tabular}{l@{\quad}l@{\quad}r@{\quad}r@{\quad}r@{\quad}r@{\quad}r@{\quad}r@{\quad}r}
  \hline
 effect &  & min & 1st qu & median & mean & 3rd qu & max & sd \\ 
  \hline
rep & par & 9.38 & 13.40 & 19.25 & 34.86 & 66.38 & 86.73 & \bf 27.00 \\ 
  rep & se & 0.61 & 1.30 & 2.91 & \bf 6.42 & 14.49 & 18.18 & 6.40 \\ 
  rep & z & 3.44 & 5.02 & 7.08 & 8.16 & 11.07 & 15.33 & 3.53 \\ 
  \hline
  pop & par & 1.20 & 1.27 & 1.29 & 1.29 & 1.32 & 1.39 & \bf 0.04 \\ 
  pop & se & 0.04 & 0.04 & 0.04 & \bf 0.04 & 0.04 & 0.04 & 0.00 \\ 
  pop & z & 32.99 & 34.71 & 35.16 & 35.17 & 35.63 & 36.73 & 0.78 \\ 
  \hline
  act & par & 1.39 & 1.47 & 1.50 & 1.49 & 1.52 & 1.57 & \bf 0.04 \\ 
  act & se & 0.03 & 0.03 & 0.03 & \bf 0.03 & 0.03 & 0.03 & 0.00 \\ 
  act & z & 44.83 & 46.66 & 47.54 & 47.40 & 48.14 & 50.08 & 1.00 \\ 
  \hline
  4cy & par & 0.47 & 0.68 & 0.79 & 0.79 & 0.87 & 1.24 & \bf 0.15 \\ 
  4cy & se & 0.08 & 0.09 & 0.10 & \bf 0.10 & 0.10 & 0.12 & 0.01 \\ 
  4cy & z & 5.37 & 7.17 & 7.94 & 7.94 & 8.54 & 10.38 & 1.07 \\ 
  \hline
  asr & par & -0.25 & -0.18 & -0.16 & -0.16 & -0.14 & -0.10 & \bf 0.03 \\ 
  asr & se & 0.02 & 0.02 & 0.03 & \bf 0.03 & 0.03 & 0.03 & 0.00 \\ 
  asr & z & -9.53 & -7.37 & -6.36 & -6.36 & -5.37 & -3.57 & 1.39 \\ 
   \hline
\end{tabular}
\end{table}

It is encouraging that all estimations over 100 samples confirm the qualitative findings from Table~\ref{tab:single}. The parameters for the repetition, popularity, activity, and four-cycle effects are positive in all 100 samples and the parameters for the assortativity effect are negative in all samples. These findings are also highly significant in all samples. The minimum absolute value of a z-value (that is, the parameter divided by its standard error) over all 100 samples and all five effects is $3.44$ (for repetition) -- still pointing to an effect that is significant at the $0.1\%$ level. Thus, the conclusions about the effects in the Wikipedia collaboration network could have been obtained by analyzing any of the 100 samples and coincide with the effects discussed in Sect.~\ref{sec:results_single}.

Differences in the reliability of estimates for the different effects, however, become apparent once we focus on the \emph{size} of parameters. Variability of parameters over the different samples shown in Table~\ref{tab:summary_fix} (also compare Fig.~\ref{fig:boxplots_fix}) varies between effects. The parameters for the popularity and activity effects -- and to a lesser extent for four-cycle and assortativity -- show relatively small variations as do their associated standard errors and z-values. For instance, for the popularity effect the interquartile range of parameters is $[1.27,1.32]$ and the minimum and maximum parameters over 100 samples span the interval $[1.20,1.39]$. Thus, had we analyzed just one of the samples, then with 50\% probability we would have obtained a parameter value from the interval $[1.27,1.32]$, with 25\% probability a value from $[1.20,1.27]$, and with 25\% probability a value from $[1.32,1.39]$. Conclusions drawn from any of these results would hardly vary in a remarkable way. Parameters for the activity effect show about the same variability and the variability of the parameters for the assortativity and four-cycle effects are somewhat higher but would still not affect conclusions in any meaningful way. By far the highest variation can be found for the repetition effect where parameters range from $9.38$ to $86.73$ with an interquartile range of $[13.40,66.38]$. We also find that the mean over the repetition parameter is considerably higher than the median pointing to a right-skewed distribution (also compare Fig.~\ref{fig:boxplots_fix}). Standard errors and z-values for the repetition effect also show high variations. Thus, while the qualitative finding of a significantly positive repetition effect does not depend on the specific sample, the magnitude of the parameter strongly varies over samples, at least for the given sample parameters $p_0$ and $m_0$. We will tackle in Sect.~\ref{sec:results_repetition} why exactly the repetition effect displays such a different behavior.

A further insight that we can draw from Table~\ref{tab:summary_fix} is that for those effects that have a low variation over samples (that is, popularity, activity, assortativity, and to a lesser extent the four-cycle effect) the standard deviation\footnote{In the whole paper, \emph{standard deviation} refers to the unbiased estimator of the sample standard deviation, where we divide the sum of squared differences to the mean by the number of values minus one.} of the model parameters over samples and the mean standard errors of the model parameters are quite similar. For these four effects, variation of standard errors is also rather small. Together, these findings seem to suggest that, if variation over samples is small, then the standard deviation of parameters caused by sampling can be approximated by the typical standard error -- a result that is consistent with asymptotic properties. This is a convenient result since standard errors can be estimated from a single sample. However, checking the relation between standard deviation of parameters over samples and mean standard error for the repetition effect, we find a much larger difference. The standard deviation of the repetition parameter over the 100 samples is $27.00$ -- much higher than the mean standard error ($6.42$), or the median standard error ($2.91$), and even higher than the maximum standard error over all 100 samples ($18.18$). Thus, judging variation in model parameters caused by sampling from the standard error of a single sample would be a poor approximation for the repetition effect.

\subsection{Varying the number of events}
\label{sec:results_prob}

\begin{figure}
    \includegraphics[width=0.49\textwidth]{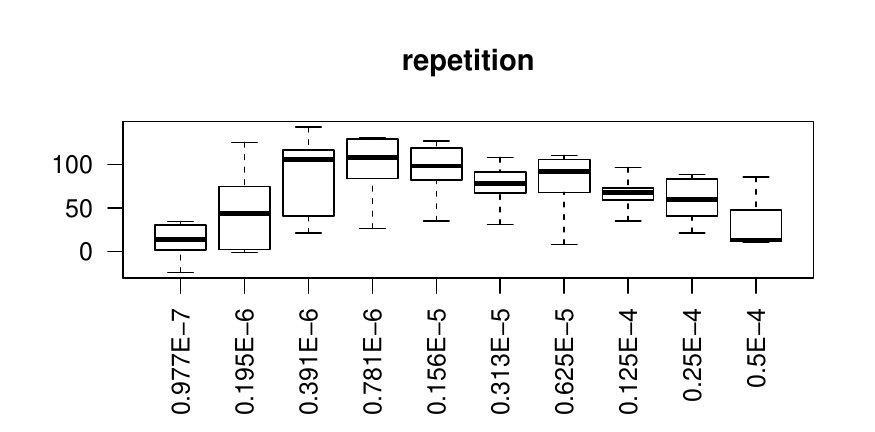}\hfill
    \phantom{\includegraphics[width=0.49\textwidth]{bp_rep_vary_prob.pdf}}
    \includegraphics[width=0.49\textwidth]{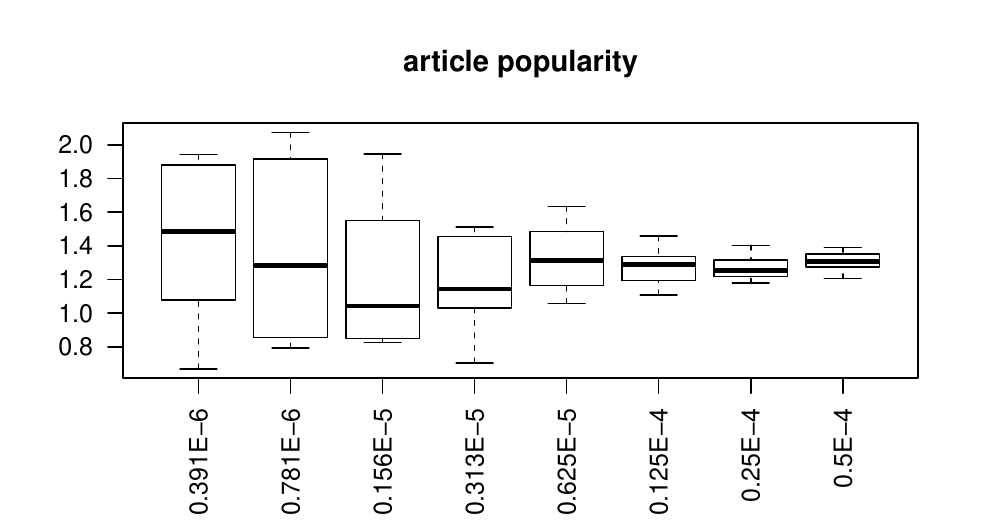}\hfill
    \includegraphics[width=0.49\textwidth]{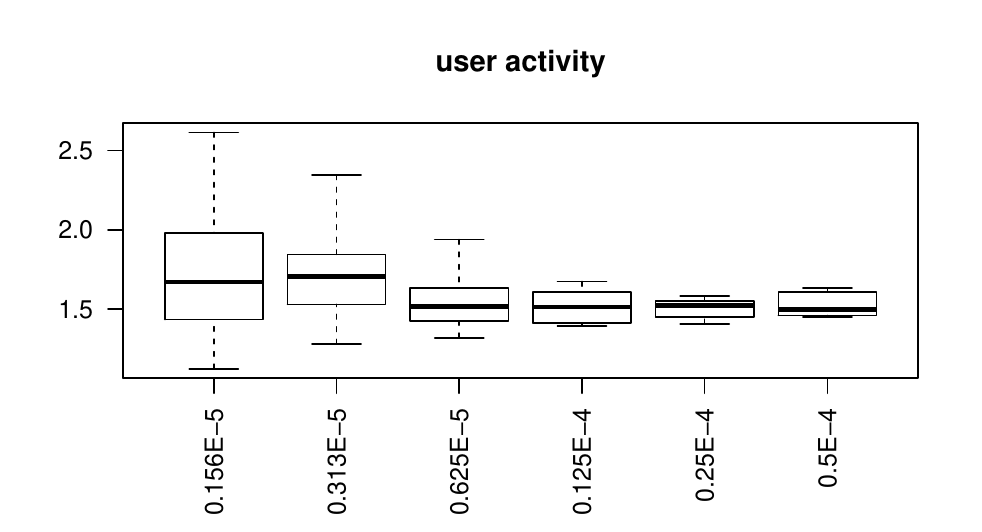}
    \includegraphics[width=0.49\textwidth]{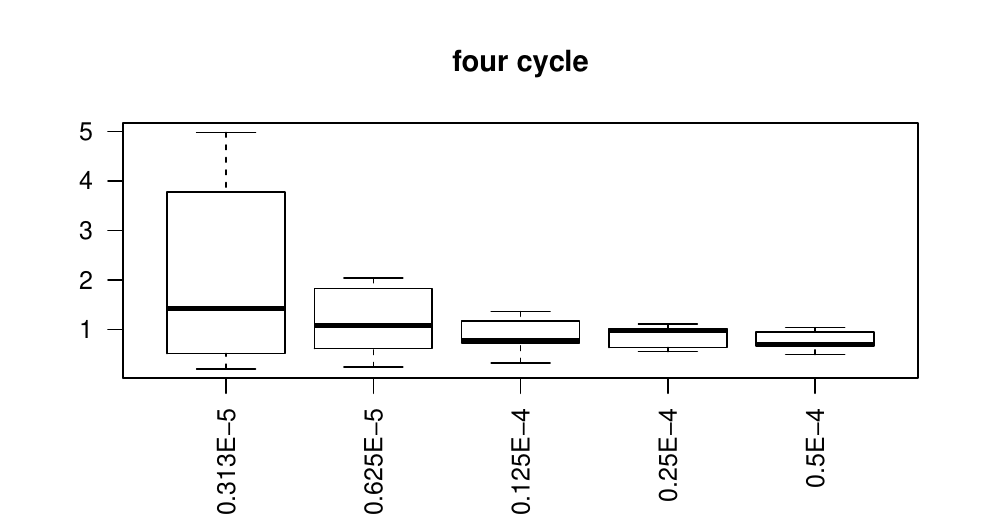}\hfill
    \includegraphics[width=0.49\textwidth]{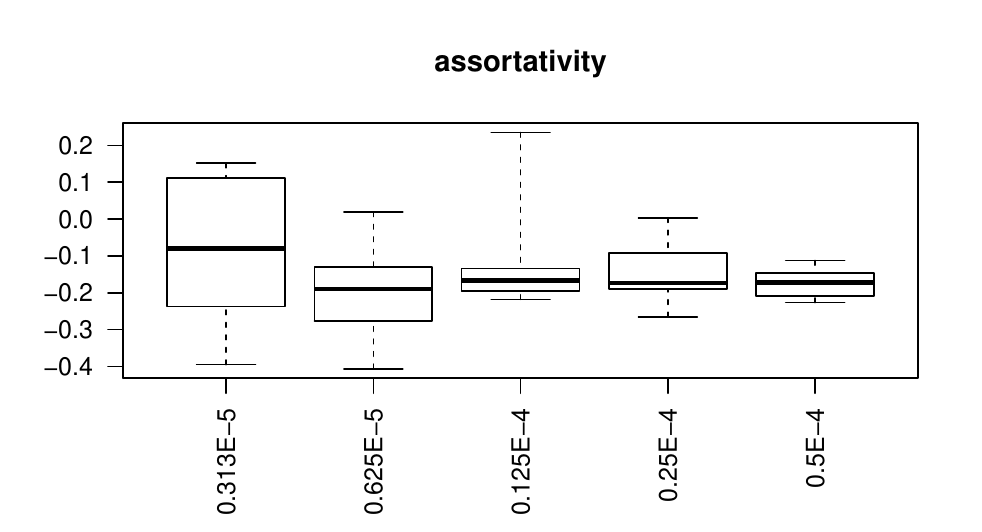}
    \caption{Boxplots of parameter estimates over 10 samples with varying $p$ (as given in the $x$-axis) and $m=5$. We discard boxplots if their dispersion causes a scaling of the $y$-axis that would make the less dispersed boxplots unreadable. (All results are given in the Appendix in Tables~\ref{tab:summary_rep_vary_prob} to~\ref{tab:summary_ass_vary_prob}.)}
    \label{fig:boxplots_vary_prob}
  \end{figure}

In a second family of experiments we fix the number of controls per event to $m=5$ and let the probability $p$ to sample from events vary from about $10^{-7}$ to $0.5\cdot 10^{-4}$ where we double $p$ in each step (compare Sect.~\ref{sec:wiki_experimental_design}). For each of these values of $p$ we draw ten samples. Model parameters for the five effects estimated for the different sample parameter settings are displayed in Figure~\ref{fig:boxplots_vary_prob}; summary statistics of estimated parameters, standard errors, and z-values for the five effects are in the Appendix in Tables~\ref{tab:summary_rep_vary_prob} to~\ref{tab:summary_ass_vary_prob}. Typically, variability of parameters decreases when the number of sampled events increases -- as it could have been expected. An exception to this rule is the repetition effect which we discuss last. 

  For the popularity effect (see Table~\ref{tab:summary_pop_vary_prob}) we find that only the two smallest sample sizes give unreliable (but also insignificant) results. We never estimated a parameter that is significantly negative at the 5\% level (corresponding to a z-value that is at least $1.96$ in absolute value). For values of $p=0.391\cdot 10^{-6}$ -- which corresponds to as little as 140 events (and five times as many controls) -- or higher we estimate only positive parameters, although not all estimates are significant. For values of $p=0.156\cdot 10^{-5}$ -- which corresponds to 561 events -- or higher we estimate only parameters for the popularity effect which are significantly positive at the 5\% level. The mean of estimated standard errors for $p=0.156\cdot 10^{-5}$ is already approximately equal to the standard deviation of parameters over the different samples. For higher values of $p$ they become even closer and both tend to zero. In summary, to reliably estimate the parameter of the popularity effect a few hundred events already seem to be sufficient.

Findings for the activity effect (see Table~\ref{tab:summary_act_vary_prob}) are similar to those of popularity. All estimated parameters for user activity are positive (though not always significant), even for the smallest sample comprising about 35 events. All parameters are significantly positive at the 5\%-level, starting from $p=0.781\cdot10^{-6}$ resulting in about 281 events and five times as many controls. The standard deviation of estimated parameters is close to the mean of the standard errors for all but the two smallest sample sizes and tends to zero when the sample size increases.

Reliable estimation of the parameter for the four-cycle effect (see Table~\ref{tab:summary_4cy_vary_prob}) requires more observations than for the degree effects popularity and activity. For a sample probability ranging from $p=0.977\cdot10^{-7}$ to $p=0.156\cdot10^{-5}$ (corresponding to about 35 to 561 events) some of the estimated four-cycle parameters are negative (in contrast to the more reliable finding reported in Table~\ref{tab:single}) -- although the negative parameters are never significant at the 5\%-level. Starting from $p=0.313\cdot10^{-5}$ all estimates are positive and for $p=0.25\cdot10^{-4}$ (corresponding to about $9,000$ sampled events), or higher, all estimated four-cycle parameters are significantly positive.

Findings for the assortativity effect (see Table~\ref{tab:summary_ass_vary_prob}) are similar to those for the four-cycle effect (keeping in mind that there is a negative assortativity effect according to the findings reported in Table~\ref{tab:single}). For $p$ up to $0.313\cdot10^{-5}$ we estimate some positive parameters for the assortativity effect -- which, however are never significant at the 5\%-level. For $p=0.5\cdot10^{-4}$ (corresponding to about $18,000$ events) all estimated assortativity parameters are significantly negative. For both effects, four-cycle and assortativity, the identity between the standard deviation of the parameters and the mean standard error is approximately valid once all parameters are significant and these values tend to zero when the sample size increases.

Findings for the repetition effect (see Table~\ref{tab:summary_rep_vary_prob}) are different. For this effect we observe that the standard deviation of the estimated parameters (that is, the uncertainty caused by sampling) does \emph{not} decrease when we increase the number of sampled events from 35 to 18,000. In contrast, the standard errors tend to decrease with increasing sample size. Together we obtain that the identity between the standard deviation of the parameters and the mean standard error does not hold for the repetition effect (at least not for this range of sample sizes). While these findings are somewhat disturbing since they imply that considerable uncertainty in the parameter sizes is caused by sampling, findings on the \emph{direction} of the repetition effect seem to be quite reliable. For all but the two smallest sample sizes we only estimate positive parameters (the negative estimates are never significant) and for $p=0.5\cdot10^{-4}$ all estimates are significantly positive. Thus, while there is considerable uncertainty about the size of the repetition effect, the danger of incorrectly rejecting the null hypothesis seems to be small. We further analyze why the repetition effect shows such an outlying behavior in Sect.~\ref{sec:results_repetition}.

\subsection{Varying the number of controls per event}
\label{sec:results_ccs}

\begin{figure}
    \includegraphics[width=0.49\textwidth]{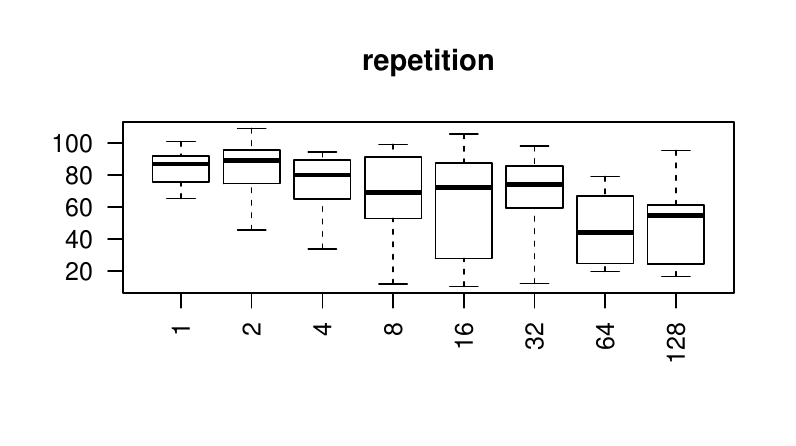}\hfill
    \phantom{\includegraphics[width=0.49\textwidth]{bp_rep_vary_ccs.pdf}}
    \includegraphics[width=0.49\textwidth]{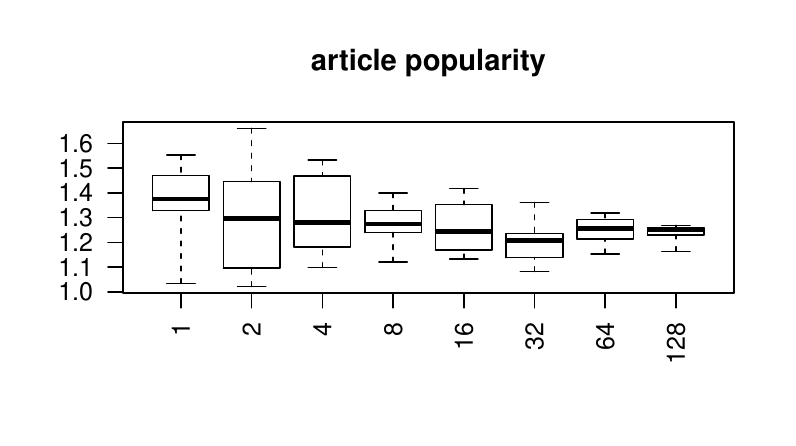}\hfill
    \includegraphics[width=0.49\textwidth]{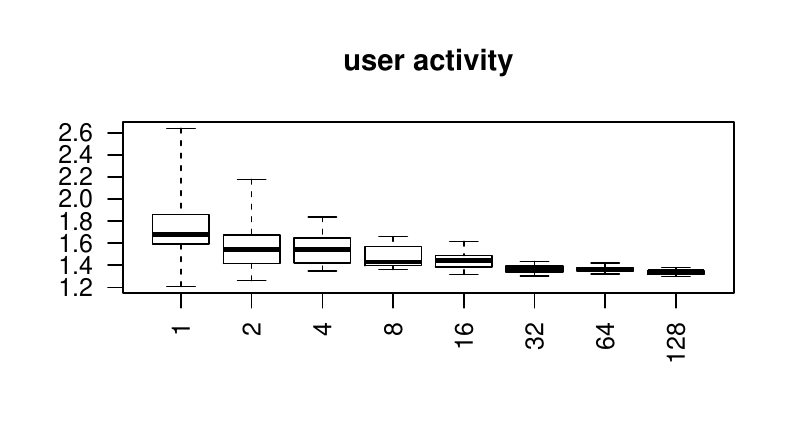}
    \includegraphics[width=0.49\textwidth]{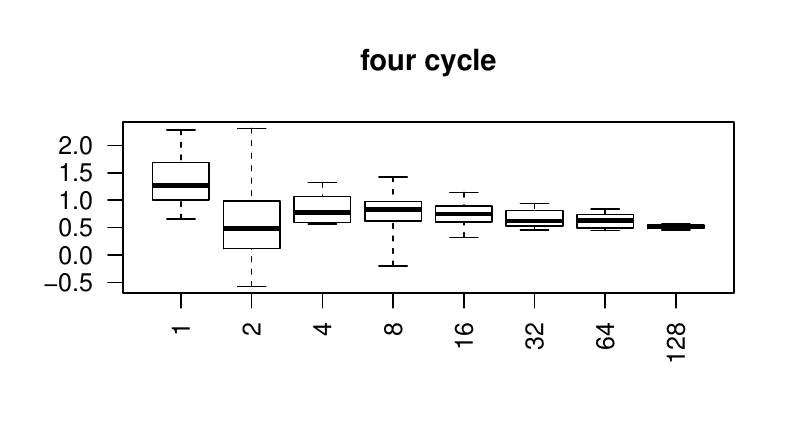}\hfill
    \includegraphics[width=0.49\textwidth]{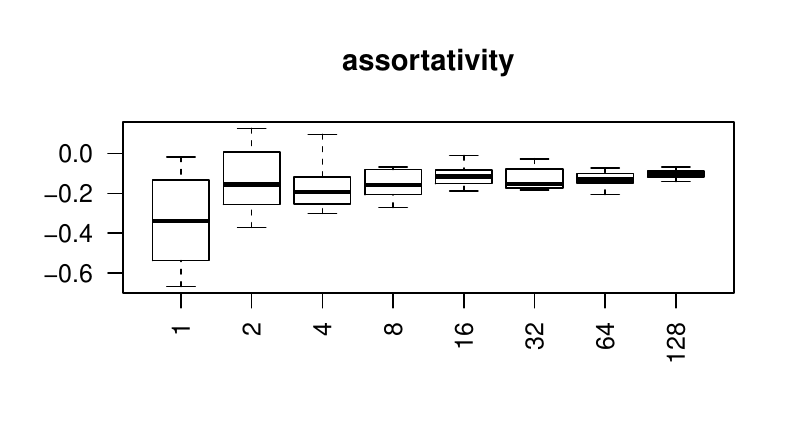}
    \caption{Boxplots of parameter estimates over 10 samples with varying $m$ (as given in the $x$-axis) and $p=1.0E-5$. (All results are given in the Appendix in Tables~\ref{tab:summary_rep_vary_ccs} to~\ref{tab:summary_ass_vary_ccs}.)}
    \label{fig:boxplots_vary_ccs}
  \end{figure}

In the next family of experiments we fix the probability to sample events to $p=10^{-5}$, corresponding to about $3,600$ sampled events, and let the number of controls per event $m$ vary from $1$ to $128$ where we double $m$ in each step (compare Sect.~\ref{sec:wiki_experimental_design}). Thus, for the smallest $m=1$ we sample about $7,200$ observations (events plus controls) and for the largest $m=128$ we sample about $464,400$ observations, which are more than twice as many observations as we analyzed in Table~\ref{tab:single} or in each of the samples considered in Sect.~\ref{sec:results_fix}, but with a much higher ratio of controls over events. For each of these values of $m$ we draw ten samples. Model parameters for the five effects estimated for the different sample parameter settings are displayed in Figure~\ref{fig:boxplots_vary_ccs}; summary statistics of estimated parameters, standard errors, and z-values for the five effects are in the Appendix in Tables~\ref{tab:summary_rep_vary_ccs} to~\ref{tab:summary_ass_vary_ccs}. In general, the variability of parameters decreases for all but the repetition effect when the number of sampled controls increases for a fixed number of sampled events.

Similar to the findings obtained by varying the number of sampled events, for the repetition effect (see Table~\ref{tab:summary_rep_vary_ccs}) we observe that the variability caused by sampling (that is, the standard deviation of parameter estimates) does \emph{not} decrease when increasing the number of observations. On the other hand, all parameter estimates are consistently positive and all estimates are significantly positive at the 5\%-level for $m=128$. In contrast to the standard deviation of the parameters, the mean standard errors do decrease with growing number of observations. Thus, for higher $m$ more and more parameters are significantly different from zero. We can observe however, that the strategy to sample more controls per event needs higher numbers of observations to achieve the same standard errors. For instance, we need more than $464,000$ observations, $3,600$ of which are events, to bring down the standard errors to about $12.9$, while $18,000$ events ($p=0.5\cdot10^{-4}$) together with $90,000$ controls yield a mean standard error of about $7.7$ (see Table~\ref{tab:summary_rep_vary_prob}). This supports the rationale of case-control sampling that ``\emph{the contribution of the nonfailures \textup{[controls]}, in terms of the statistical power of the study, will be negligible compared to that of the failures \textup{[events]}}'' \citep{bgl-mascdcphm-95}. In Sect.~\ref{sec:results_co_vary} we will analyze more systematically whether there is an optimal number of controls per event.

For the popularity effect (see Table~\ref{tab:summary_pop_vary_ccs}) we find that not only the mean standard error, but also the standard deviation of the parameters gets smaller with growing number of observations for a fixed number of events. All parameter estimates for this effect are significantly positive, for all values of $m$ that we considered in this type of experiments. We find that increasing the number of controls for a fixed number of events seems to require more observations to bring down the standard errors -- or the standard deviation of parameters -- than when sampling more events but fewer controls per event. For instance, the setting with sample parameters $p=10^{-5}$ and $m=64$ yields about $234,000$ observations and achieves about the same standard deviation of the parameters as the setting with $p=0.5\cdot10^{-4}$ and $m=5$ which yields about $108,000$ observations. Findings for the activity effect (see Table~\ref{tab:summary_act_vary_ccs}) are very similar to those for the popularity effect.

Estimating the parameters of the four-cycle effect (see Table~\ref{tab:summary_4cy_vary_ccs}) requires more observations than for the degree effects. The setting with $p=10^{-5}$ and $m=2$ is actually the only one in which we estimate a parameter that is significant at the $5\%$-level and that points in the opposite direction than the corresponding parameter given in Table~\ref{tab:single}. Considering that we draw and analyze in total 360 samples, it seems to be acceptable to incorrectly reject a null hypothesis once at the 5\%-level. We find in Table~\ref{tab:summary_4cy_vary_ccs} that starting from $m=16$ we only estimate significantly positive four-cycle parameters.

Estimated parameters for the assortativity effect (see Table~\ref{tab:summary_ass_vary_ccs}) are in most cases negative (as in Table~\ref{tab:single}), the positive estimates are never significant, and starting from $m=64$ we only estimate significantly negative parameters.

\subsection{Varying the ratio of controls per event for a given budget of observations}
\label{sec:results_co_vary}

\begin{figure}
    \includegraphics[width=0.49\textwidth]{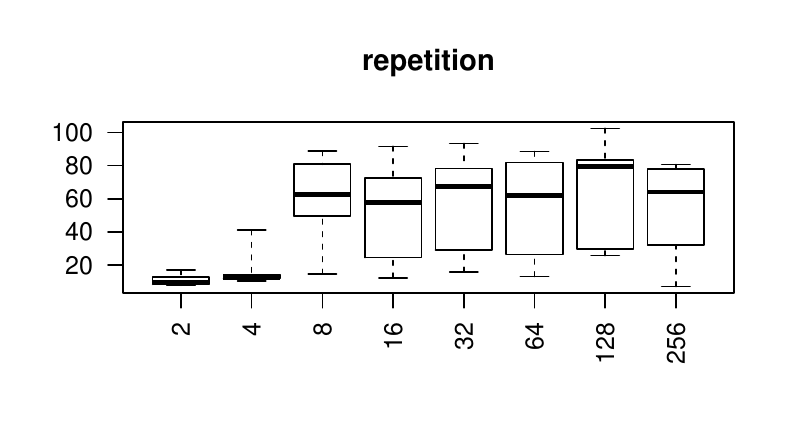}\hfill
    \phantom{\includegraphics[width=0.49\textwidth]{bp_rep_co_vary.pdf}}
    \includegraphics[width=0.49\textwidth]{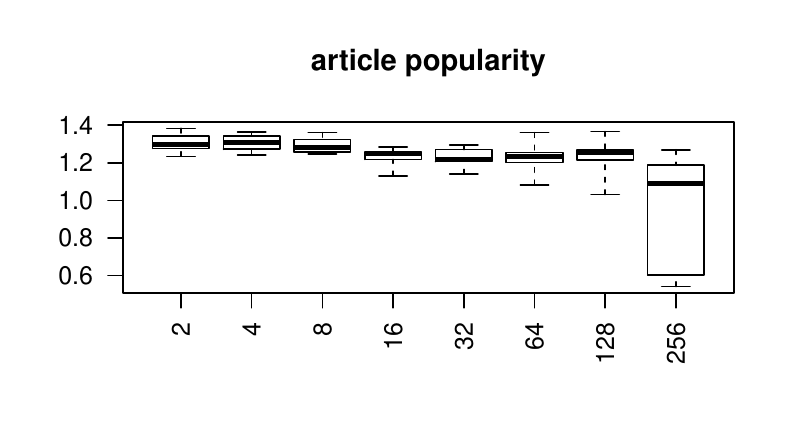}\hfill
    \includegraphics[width=0.49\textwidth]{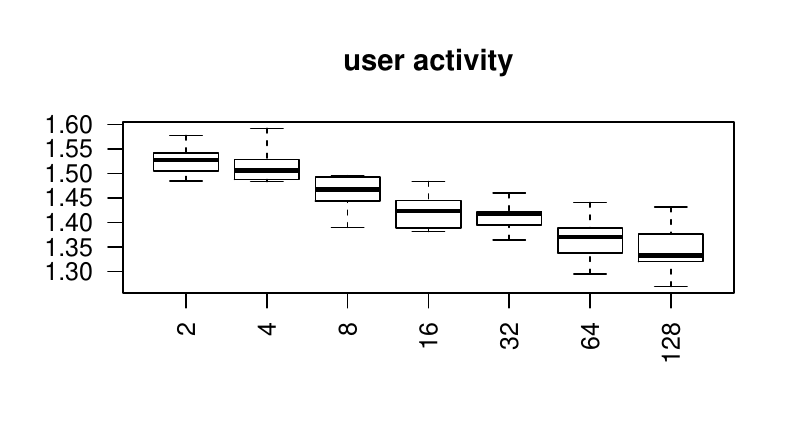}
    \includegraphics[width=0.49\textwidth]{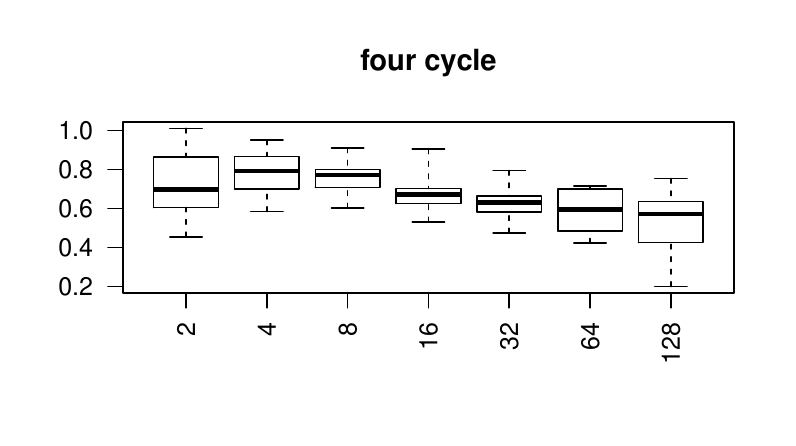}\hfill
    \includegraphics[width=0.49\textwidth]{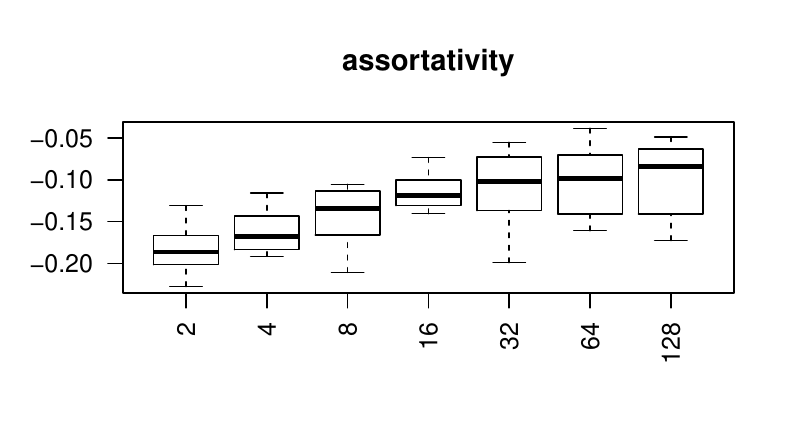}
    \caption{Boxplots of parameter estimates over 10 samples with varying $m$ (as given in the $x$-axis) and varying $p$ satisfying $(m+1)\cdot p=6\cdot 0.0001$. We discard boxplots if their dispersion causes a scaling of the $y$-axis that would make the less dispersed boxplots unreadable. (All results are given in the Appendix in Tables~\ref{tab:summary_rep_co_vary} to~\ref{tab:summary_ass_co_vary}.)}
    \label{fig:boxplots_co_vary}
  \end{figure}

In the final set of computational experiments we explore whether, given a limited budget to allocate to observations, it is preferable to draw fewer events and more controls per events or whether more events and fewer controls yield more reliable estimates. Thus, we fix the number of observations to about $216,000$ -- the number of observations resulting from the initial sample parameters $p=10^{-4}$ and $m=5$ considered in Sect.~\ref{sec:results_fix}. We let the number of controls per event $m$ vary from $2$ to $256$ where we double $m$ in each step (compare Sect.~\ref{sec:wiki_experimental_design}). For a given $m$ we set the probability to sample events to $p=6\cdot10^{-4}/(m+1)$ to keep the number of sampled observations constant. Thus, for the smallest $m=2$ we sample about $72,000$ events and $144,000$ controls and for the largest $m=256$ we sample about 840 events and about $215,000$ controls. For each of these values of $m$ and $p$ we draw ten samples. Model parameters for the five effects estimated for the different sample parameter settings are displayed in Figure~\ref{fig:boxplots_co_vary}; summary statistics of estimated parameters, standard errors, and z-values for the five effects are in the Appendix in Tables~\ref{tab:summary_rep_co_vary} to~\ref{tab:summary_ass_co_vary}. In general, the variability of parameters increases when we sample additional controls at the expense of sampling fewer events, so that sampling more events and fewer controls seems to be the better investment.

Findings for the repetition effect (see Table~\ref{tab:summary_rep_co_vary}) fit into the general pattern that variability of estimates increases if the number of sampled events decreases and the number of controls increases accordingly. For this effect, a considerable jump becomes apparent when $m$ changes from 2 over 4 to 8. For $m=2$ the standard deviation of the parameters is $3.33$ and the mean standard error is equal to $0.97$. These numbers increase to a standard deviation of $9.09$ and a mean standard error of $1.87$ for $m=4$ and to a standard deviation of $26.06$ and a mean standard error of $13.34$ for $m=8$. For higher values of $m$, the standard deviation is consistently above $20$ and the mean standard error consistently above $10$. Starting from $m=64$, some of the estimated parameters are no longer significant at the 5\%-level, although all estimates are positive. 

These findings seem to suggest that increasing the ratio of events over controls, while keeping the number of observations constant, is a strategy for reliably estimating the repetition parameter. As a cautionary note we want to indicate already here that results from Sect.~\ref{sec:results_repetition} suggest that this is misleading: the repetition variable is very skewed and almost all controls have the same value (equal to zero). We discuss estimation of the repetition effect in more detail in Sect.~\ref{sec:results_repetition}.

Parameter estimates for the popularity effect (see Table~\ref{tab:summary_pop_co_vary}) are almost unaffected by the ratio of controls over events for values from $m=2$ to $m=32$. Standard errors and standard deviation of the parameter estimates increase slightly for $m=64$ and $m=128$ and increase sharply for $m=256$. For this last value we also get the first non-significant estimates, although all estimates are positive. Findings for the activity effect (see Table~\ref{tab:summary_act_co_vary}) and the four-cycle effect (see Table~\ref{tab:summary_4cy_co_vary}) are similar to those for the popularity effect, showing a sharp increase in variability for $m=256$ -- although all estimated parameters for activity and four-cycle remain significant for the analyzed range of $m$.

For the assortativity effect (see Table~\ref{tab:summary_ass_co_vary}) we make the same observation that variability increases sharply for $m=256$. Some estimates for this parameter become insignificant for $m=32$ or higher, although all estimates are negative.

In general, it seems to be a more useful investment to sample events, rather than controls, so that the number of controls per event is best set to a small number. For values in the range of $m=2$ to $m=16$ the actual value seems to make little difference. The remarkable differences between $m=2$, 4, or 8 for the repetition effect rather seems to be a spurious finding, as discussed in Sect.~\ref{sec:results_repetition}.

\subsection{Examining the repetition effect}
\label{sec:results_repetition}

A common pattern in the results discussed above is that the estimated parameters of the repetition effect had a much higher variability across samples than those of the other effects. Disturbingly -- and in contrast to findings for the other effects in the model -- the variability of the repetition parameter does often not decrease when we increase the sample size, at least not for the ranges of sample parameters considered so far. We have also found -- again in contrast to findings for the other effects -- that the standard errors of the repetition effect are a poor approximation for the variability caused by sampling. Thus, while the \emph{direction} of this effect is consistently positive, its \emph{magnitude} can only be estimated with a relatively high degree of uncertainty.

In this section we perform an additional analysis to find explanations why the estimated parameters of the repetition effect have such a high standard deviation, compared to the other effects. We analyze the combined 100 samples obtained with $p=10^{-4}$ and $m=5$, that is, the aggregation of the samples used in Sect.~\ref{sec:results_fix}. These constitute more than 21 million observations, 16.7\% of which are events and 83.3\% are controls. The densities, that is, the proportions of observations with non-zero values for the five statistics are reported in Table~\ref{tab:densities}. We compute and report densities separately for all observations, restricted to events, and restricted to controls (i.\,e., dyads not experiencing an event at the given time).

\begin{table}
  \caption{Density (i.\,e., proportion of observations with non-zero values) for the five statistics, reported for all observations (2nd column), restricted to events (3rd column), and restricted to controls (4th column) computed from the about 21 million observations obtained by aggregating the 100 samples with $p=10^{-4}$ and $m=5$, used in Sect.~\ref{sec:results_fix}.}
  \label{tab:densities}
\centering\begin{tabular}{l@{\quad}r@{\quad}r@{\quad}r}
  \hline
statistic & density.obs & density.events & density.controls \\ 
  \hline
repetition & 0.084385 & 0.506282 & \bf 0.000006 \\ 
  popularity & 0.780180 & 0.937911 & 0.748634 \\ 
  activity & 0.378371 & 0.977290 & 0.258588 \\ 
  four.cycle & 0.163034 & 0.766557 & 0.042329 \\ 
  assortativity & 0.332671 & 0.917340 & 0.215737 \\ 
   \hline
\end{tabular}
\end{table}

In short, the reason for the anomalous behavior of the repetition parameter seems to be the extremely low number of controls that have a non-zero value in the repetition statistic. The density of the repetition statistic restricted to controls is just about $6\cdot 10^{-6}$. In the aggregation of all 100 samples, we have about 100 controls with non-zero repetition values. Thus, each of the samples analyzed in Sect.~\ref{sec:results_fix} has on average just one control with a non-zero repetition value. The high variability of estimated repetition parameters across samples seems to be caused by the randomness of sampling none, just one, or several controls with a non-zero value in the repetition statistic. 

We can now also better understand the result reported in Sect.~\ref{sec:results_co_vary} (also see Table~\ref{tab:summary_rep_co_vary}) that sampling fewer controls (while increasing the number of events accordingly) seems to bring down the variability of the repetition parameter to a relatively low value. Decreasing the number of sampled controls actually also decreases the probability that the sample includes an exceptional non-event dyad with non-zero repetition variable. For low numbers of controls it becomes unlikely to sample even a single one of these exceptional controls. In such situations, the parameter estimates become only seemingly reliable, since the \emph{sample} variation of the explanatory variable among the controls suddenly drops to zero. Clearly this is misleading: if our sample accidentally included one of the exceptional controls with non-zero repetition variable, then the parameter estimate would change strongly.

The other four statistics are much more homogeneously distributed. The statistic with the next lowest density is the four-cycle statistic, which has a density of $0.04$ among the controls. This translates to more than $760,000$ controls with a non-zero four-cycle statistic in the aggregation of all 100 samples and on average more than $7,600$ in each of the samples  analyzed in Sect.~\ref{sec:results_fix}. Apparently this number is sufficient to estimate parameters with low uncertainty caused by sampling. The other statistics, popularity, activity, and assortativity, are more evenly distributed than the four-cycle statistic. We discuss in Sect.~\ref{sec:strata} a possible way to reduce the variability of the repetition parameter.

We emphasize that it is important to check the density of non-zero values -- or, more generally, the distributions of explanatory variables -- separately for events and controls. Table~\ref{tab:densities} reveals that the density of the repetition statistic over all observations (events and controls) is not exceptional.

\section{Summary of results and future work}
\label{sec:discussion}

In general, findings reported in this paper are encouraging. Reliably estimating relational event models on event networks with millions of nodes and hundreds of millions of events seems to be possible by sampling some tens of thousands of events and a small number of controls per event. For some of the analyzed effects -- most notably the degree effects \emph{popularity} and \emph{activity} -- an even smaller sample size comprising a few hundred events seems to be sufficient. In general, z-values (that is, parameters divided by their standard errors) seem to provide a reliable criterion for whether estimated parameters are significantly positive or negative. This suggests that one way to scale up REMs is by applying relatively simple and well-established sampling schemes.

Thus, the \emph{direction} of effects can apparently be estimated in a reliable way with relatively small sample sizes. A different task however is to assess the \emph{size} of parameters with small variability. In this aspect, the repetition effect turned out to be exceptionally difficult.

For those settings of the sample parameters that yield a low parameter variation caused by sampling, the standard deviation of parameters across samples is approximately equal to the mean standard errors, as suggested by asymptotic results. This is a convenient result since standard errors can be estimated from a single sample. Our results related with the repetition parameter, however, indicate that this approximate identity can not be taken for granted: the standard deviation of the repetition parameter is consistently much higher than its typical standard error. We identified the extreme sparseness of the repetition variable among the controls as a likely reason for this finding. Only six controls in a million have a non-zero value in the repetition variable. Reliable estimation seems to need more than just a few of these exceptional controls, implying that the sampled set of controls should have several millions of elements (see, however, Sect.~\ref{sec:strata} for a more efficient way to get reliable estimates).

Due to the above findings, analysts should not just estimate parameters and standard errors for a single sample and draw their conclusions from the resulting z-values. It is definitely recommendable to check the distribution of statistics separately for sampled events and for sampled controls. It seems to be problematic if any of these distributions reveal that almost all controls, or almost all events, have the same value. Just checking the distribution of statistics over \emph{all} observations (events and controls) is insufficient to detect outlying statistics -- as is the variance, covariance, or correlation of statistics computed over \emph{all} observations, see Table~\ref{tab:cov_cor}. Additionally to checking empirical distributions one could also infer by theoretical arguments whether certain statistics are likely to be heavily skewed. In our concrete model we could conclude by comparing the size of the risk set (more than 30 trillion) with the number of dyadic events (about 360 million) that the repetition variable cannot be evenly distributed over the risk set. (On the other hand, it seems to be much more difficult to conclude from theoretical considerations why the four-cycle statistic has a relatively high density, equal to $0.04$, among the controls.) In situations in which almost all controls take a single value in some statistic, the most efficient way seems to change the sampling design as it is sketched in Sect.~\ref{sec:strata}.

If computational resources are sufficient, analysts can draw several samples with the same sample parameters, fit models separately to these, and then assess the variation caused by sampling (as we do in this paper). If for computational reasons (or otherwise) repeated sampling is not possible, bootstrap sampling \citep{efron1992bootstrap} might be an alternative to assess variation of parameters from one given set of observations. (We experimentally applied bootstrap sampling to one single sample and found parameter variation estimated from bootstrapping to be very similar to the one obtained in Sect.~\ref{sec:results_fix}.) However, in light of the findings discussed in Sects.~\ref{sec:results_co_vary} and~\ref{sec:results_repetition}, re-sampling (or bootstrap sampling) does not render it unnecessary to check distributions of statistics: if some statistic is extremely degenerate, even large combined sample sizes might miss the exceptional observations and would incorrectly suggest a small parameter variation. Inspecting the values of statistics separately on sampled events and sampled controls might reveal that, for instance, all sampled controls take the same value, and thus could point to near-degenerate distributions.

We also analyzed the trade-off between sampling more events and fewer controls per event \emph{versus} sampling fewer events and more controls per event, given a limited budget of observations. Results seem to confirm that events are more valuable than controls \citep{bgl-mascdcphm-95} so that $m$, the number of controls per event is best kept at a small number -- provided that the number of observed events to sample from is sufficiently high. In our experiments, the actual value of $m$ seemed to make little difference for values up to $m=16$, but choosing higher values of $m$ -- at the expense of sampling fewer events -- increased parameter variation.

We fitted model parameters also after standardizing explanatory variables (these estimates are not reported in this paper). That is, we transformed all the statistics by subtracting their mean and dividing by their standard deviation before fitting the model. In general, standardization resulted in lower variation of the estimated model parameters; this applies especially to the repetition effect, as variation for the other effects is small anyway. However, dividing statistics by their standard deviation has a side effect when we want to compare estimations on samples with varying numbers of controls per event. Typically, controls have much smaller values in the various statistics. Thus, if we include more controls per event we typically get lower standard deviations of statistics. Due to these data characteristics, standardization of statistics resulted in parameter estimates that systematically tend to zero with higher ratio of controls per event (although standardization does not affect z-values, since standard errors are scaled with the same factor as parameters). In this paper, we leave it open whether standardization of statistics is recommendable in general. In our study we do not want to scale statistics with different factors in different samples -- for this reason we report results obtained without standardization. Yet, in other studies standardization might be appropriate to improve comparability of parameter sizes over different effects.

\subsection{Improving estimation of the repetition parameter: stratified sampling}
\label{sec:strata}

Reliable estimation of the \emph{size} of the repetition parameter turned out to be much more difficult than for the other effects. We have found in Sect.~\ref{sec:results_repetition} that the extreme sparsity of the repetition statistic \emph{among the controls} is a likely cause for this. One obvious way to bring down the sample variance of the repetition parameter is to further increase the sample size, until the samples include a sufficiently high number of controls with non-zero values. However, this approach is inefficient. A more efficient approach seems to be stratified sampling \citep{langholz1995counter} where we divide observations into strata -- for instance, those dyads that have zero weights, compare Sect.~\ref{sec:wiki_model}, and those that have non-zero weights -- and sample controls evenly from these strata.

When applying stratified sampling we have to re-weight observations when estimating the Cox-proportional hazard model. Stratification typically leads to estimates with lower variation \citep{langholz1995counter}; it has been applied to fitting REM in \citet{vpr-remslm-15}. We do not consider stratified sampling in this paper, but suggest it as a strategy to estimate REM in situations with near-degenerate explanatory variables, that is, variables that are (almost) constant on the vast majority of instances, but take considerably different values on few instances.

\subsection{Beyond the Cox proportional hazard model}
\label{sec:beyond_cox}

In this paper we considered the Cox proportional hazard model where the estimated parameters explain which statistics increase or decrease the rate of events but where the baseline hazard $\lambda_0(t)$, see Eq.~(\ref{eq:lambda}), is left unspecified. This model, which corresponds to the ``ordinal model'' from \citet{b-refsa-08}, is appropriate in many situations since we are often less interested in the absolute rate of events but rather in the factors that increase or decrease the likelihood of events. In our setting, the baseline hazard $\lambda_0(t)$ corresponds to the rate of edit events at time $t$ in the whole of Wikipedia, that is, independent of the active user or the target article. This global rate of events changes systematically over the lifetime of Wikipedia and also changes periodically by the day of the week and the time of the day \citep{yasseri2012circadian}. Thus, it seems to be even preferable to filter out these regularities in the event rate -- as it is done in the Cox proportional hazard model -- if the goal of a study is to analyze network effects explaining dyadic event rates. 

Yet, in other applications of relational event models analysts might be interested in the absolute event rate, or from another point of view, in the expected time to events, and not just in whether the rate on some dyads tends to be higher than on others. Relational event models that achieve this are well-known \citep{b-refsa-08} -- but applying case-control sampling to these REMs that take time information from a ratio scale, rather than an ordinal scale, brings up some additional issues. It is easy to see that manipulating the number of controls per event -- as it is done by case-control sampling -- affects at least the intercept of a parametric specification of the full hazard function. (In contrast, a Cox proportional hazard model has no intercept since an intercept would be absorbed by the baseline hazard.) 

An alternative for analysts interested in the absolute hazard rate would be to estimate the integrated baseline hazard $\Lambda_0(t)=\int_0^t\lambda_0(t')\,dt'$, where $\lambda_0(t)$ is the baseline hazard from Eq.~(\ref{eq:lambda}), in a non-parametric way -- an approach that can be combined with case-control sampling \citep{bgl-mascdcphm-95}. If we also sample from the observed events with probability $p<1$, additionally to case-control sampling, we just have to scale the obtained baseline hazard with $1/p$.

\subsection{Reproducibility}
\label{sec:reproducibility}

Computation of explanatory variables for the experiments presented in this paper has been done with the open-source software \emph{eventnet}\footnote{\url{https://github.com/juergenlerner/eventnet}} and parameter estimation has been done with the \texttt{survival} package\footnote{\url{https://CRAN.R-project.org/package=survival}} \citep{therneau2013modeling} in the R software for statistical computing. The preprocessed list of 360 million relational events which constitutes our input data is publicly available at \url{https://doi.org/10.5281/zenodo.1626323}. The eventnet configuration, and an R script to fit models are also available online.\footnote{\url{https://github.com/juergenlerner/eventnet/wiki/Large-event-networks-(tutorial)}} This facilitates to reproduce experiments as the ones presented in this paper -- potentially varying sample parameters, model statistics, model estimation procedure, or the empirical data used to fit the model.

\section{Conclusion}
\label{sec:conclusion}

Relational event networks naturally result from computer-mediated communication and collaboration or automated data collection strategies that record social interaction. Networks obtained in this way are often large. Relational event models provide an appropriate statistical framework to model event networks without having to aggregate dyadic events over time intervals. Yet, studies that apply REM to large event networks are surprisingly rare.

Previous work \citep{vpr-remslm-15} proposed case-control sampling to reduce the size of the risk set which is often quadratic in the number of nodes. In this paper we systematically assess the variation in parameter estimates that have been obtained by a combination of case-control sampling and uniform sampling from the observed events.

Results from this paper give an encouraging message. It seems to be possible to reliably fit relational event models to networks with millions of nodes and hundreds of millions of events by analyzing a sample of some tens of thousands of events and a small number of controls per event. Some network effects, such as degree effects, seem to require even much smaller sample sizes comprising a few hundred events to be reliably estimated. We identified characteristics of the data that might cause a relatively high variability in the estimation of some effects and recommend that analysts should check distributions of explanatory variables -- separately for sampled events and controls -- to recognize such situations.

Future work might experimentally assess empirical properties of different sampling strategies, including stratified sampling \citep{vpr-remslm-15} which might be a way to overcome situations of very skewed variables. Additionally to case-control sampling and sampling observed events used for fitting the model, one might also consider sampling from the input events that define the network of past events, see Sect.~\ref{sec:wiki_model}. Such an approach might be a way to analyze even larger event networks -- especially if the unrestricted network of past events is too large to be kept in computer memory. In this paper we maintained the network of past events as a function of all input events without sampling, since this ensures that explanatory variables are computed correctly. Sampling from the input events would require to conduct sensitivity experiments similar to the ones presented in this paper. (We note that the scenario of using only the sampled events to build up the network of past events is equivalent to settings with missing data -- given the assumption that dyadic observations are missing completely at random. In our situation, where we have the full data available and sample only to reduce computational runtime, the assumtion of ``missing completely at random'' is guaranteed by design.) Last but not least, computation times of statistics could potentially be reduced by improvements in graph algorithms. For instance, in our model the computation of the four-cycle statistic has an asymptotic runtime which is by orders of magnitude larger than the runtime needed to compute the other statistics. Algorithms for more efficiently maintaining counts of four-cycles in dynamic networks exist \citep{eppstein2012extended} but do not match the usage scenario of this paper where we have many updates and need four-cycle counts only for a much smaller number of dyads. 

\paragraph{Conflicts of interest:} the authors have nothing to disclose.


\appendix

\section{Detailed results: distributions of parameters, standard errors, and z-values}

\begin{table}
  \caption{Summary statistics of parameters (\emph{par}), standard errors (\emph{se}), and z-values (\emph{z}) for the \textbf{\emph{repetition}} effect over 10 samples with $m=5$ and varying $p$ (as given in the first column).}
  \label{tab:summary_rep_vary_prob}
\centering\begin{tabular}{l@{\quad}l@{\quad}r@{\quad}r@{\quad}r@{\quad}r@{\quad}r@{\quad}r@{\quad}r}
  \hline
 $p$ &  & min & 1st qu & median & mean & 3rd qu & max & sd \\ 
  \hline
0.977E-7 & par & -24.29 & 2.71 & 13.81 & 13.41 & 29.71 & 34.48 & \bf 18.40 \\ 
  0.977E-7 & se & 141.56 & 10145.18 & 12967.66 & \bf 12726.78 & 14125.86 & 30204.59 & 7990.80 \\ 
  0.977E-7 & z & -0.00 & 0.00 & 0.00 & 0.00 & 0.00 & 0.01 & 0.00 \\ 
   \hline
  0.195E-6 & par & -1.20 & 8.73 & 43.78 & 49.91 & 74.09 & 125.75 & \bf 45.21 \\ 
  0.195E-6 & se & 101.80 & 304.21 & 3534.09 & \bf 7374.75 & 8400.51 & 37400.83 & 11410.16 \\ 
  0.195E-6 & z & -0.00 & 0.00 & 0.01 & 0.09 & 0.19 & 0.28 & 0.11 \\ 
   \hline
  0.391E-6 & par & 20.89 & 43.76 & 106.11 & 85.64 & 115.40 & 143.17 & \bf 45.61 \\ 
  0.391E-6 & se & 70.17 & 730.62 & 2352.02 & \bf 4732.95 & 10076.41 & 10852.35 & 4794.37 \\ 
  0.391E-6 & z & 0.00 & 0.01 & 0.04 & 0.19 & 0.08 & 1.50 & 0.46 \\ 
   \hline
  0.781E-6 & par & 26.31 & 85.47 & 108.36 & 97.35 & 128.36 & 131.21 & \bf 38.30 \\ 
  0.781E-6 & se & 181.24 & 902.81 & 1960.73 & \bf 3318.56 & 5816.34 & 9886.83 & 3350.04 \\ 
  0.781E-6 & z & 0.01 & 0.01 & 0.04 & 0.13 & 0.12 & 0.69 & 0.21 \\ 
   \hline
  0.156E-5 & par & 34.95 & 83.03 & 98.40 & 94.46 & 118.24 & 127.19 & \bf 29.72 \\ 
  0.156E-5 & se & 39.04 & 226.98 & 532.09 & \bf 1232.39 & 1222.64 & 5309.28 & 1659.93 \\ 
  0.156E-5 & z & 0.02 & 0.08 & 0.15 & 0.29 & 0.46 & 0.90 & 0.29 \\ 
   \hline
  0.313E-5 & par & 31.09 & 69.45 & 78.68 & 78.95 & 91.20 & 108.30 & \bf 22.71 \\ 
  0.313E-5 & se & 43.21 & 56.15 & 118.55 & \bf 355.64 & 186.65 & 2534.21 & 767.86 \\ 
  0.313E-5 & z & 0.04 & 0.50 & 0.65 & 0.79 & 0.79 & 2.23 & 0.61 \\ 
   \hline
  0.625E-5 & par & 8.06 & 72.90 & 92.39 & 83.28 & 105.36 & 110.79 & \bf 31.64 \\ 
  0.625E-5 & se & 1.74 & 56.08 & 91.67 & \bf 88.97 & 131.87 & 145.67 & 48.37 \\ 
  0.625E-5 & z & 0.70 & 0.75 & 0.98 & 1.41 & 1.41 & 4.64 & 1.20 \\ 
   \hline
  0.125E-4 & par & 34.91 & 59.24 & 67.53 & 67.14 & 72.52 & 96.80 & \bf 17.63 \\ 
  0.125E-4 & se & 25.11 & 31.15 & 42.43 & \bf 51.70 & 53.42 & 111.70 & 29.81 \\ 
  0.125E-4 & z & 0.87 & 1.00 & 1.39 & 1.54 & 1.88 & 2.79 & 0.66 \\ 
   \hline
  0.25E-4 & par & 21.49 & 43.75 & 59.93 & 59.73 & 80.65 & 88.57 & \bf 23.27 \\ 
  0.25E-4 & se & 7.18 & 21.93 & 31.14 & \bf 26.59 & 33.90 & 34.77 & 9.66 \\ 
  0.25E-4 & z & 1.60 & 2.06 & 2.44 & 2.31 & 2.54 & 2.99 & 0.45 \\ 
   \hline
  0.5E-4 & par & 10.33 & 11.75 & 13.30 & 31.09 & 42.50 & 85.84 & \bf 29.53 \\ 
  0.5E-4 & se & 0.97 & 1.37 & 1.92 & \bf 7.74 & 13.17 & 23.90 & 9.19 \\ 
  0.5E-4 & z & 3.19 & 3.58 & 6.98 & 6.44 & 8.62 & 10.60 & 2.76 \\ 
   \hline
\end{tabular}
\end{table}

\begin{table}
  \caption{Summary statistics of parameters (\emph{par}), standard errors (\emph{se}), and z-values (\emph{z}) for the \textbf{\emph{popularity}} effect over 10 samples with $m=5$ and varying $p$ (as given in the first column).}
  \label{tab:summary_pop_vary_prob}
\centering\begin{tabular}{l@{\quad}l@{\quad}r@{\quad}r@{\quad}r@{\quad}r@{\quad}r@{\quad}r@{\quad}r}
  \hline
$p$ &  & min & 1st qu & median & mean & 3rd qu & max & sd \\ 
  \hline
0.977E-7 & par & -3.48 & 2.75 & 11.72 & 12.52 & 20.78 & 34.24 & \bf 11.73 \\ 
  0.977E-7 & se & 0.71 & 107.06 & 3296.22 & \bf 2765.42 & 4865.10 & 5552.37 & 2440.08 \\ 
  0.977E-7 & z & -0.57 & 0.00 & 0.00 & 0.05 & 0.03 & 0.76 & 0.32 \\ 
  \hline
  0.195E-6 & par & -21.66 & 1.01 & 1.62 & 5.47 & 7.05 & 65.67 & \bf 23.47 \\ 
  0.195E-6 & se & 0.73 & 1.00 & 46.57 & \bf 2504.07 & 948.11 & 22365.06 & 6996.94 \\ 
  0.195E-6 & z & -0.34 & 0.00 & 0.07 & 0.56 & 1.26 & 1.84 & 0.85 \\ 
  \hline
  0.391E-6 & par & 0.67 & 1.16 & 1.49 & 1.46 & 1.84 & 1.94 & \bf 0.43 \\ 
  0.391E-6 & se & 0.41 & 0.55 & 0.77 & \bf 0.83 & 1.05 & 1.39 & 0.35 \\ 
  0.391E-6 & z & 1.16 & 1.46 & 1.84 & 1.90 & 2.28 & 2.92 & 0.59 \\ 
  \hline
  0.781E-6 & par & 0.79 & 0.87 & 1.28 & 1.34 & 1.79 & 2.07 & \bf 0.51 \\ 
  0.781E-6 & se & 0.32 & 0.36 & 0.49 & \bf 0.49 & 0.58 & 0.75 & 0.14 \\ 
  0.781E-6 & z & 1.61 & 2.49 & 2.66 & 2.76 & 3.06 & 3.97 & 0.64 \\ 
  \hline
  0.156E-5 & par & 0.83 & 0.87 & 1.04 & 1.23 & 1.51 & 1.95 & \bf 0.45 \\ 
  0.156E-5 & se & 0.23 & 0.28 & 0.31 & \bf 0.31 & 0.33 & 0.39 & 0.05 \\ 
  0.156E-5 & z & 2.69 & 3.19 & 3.84 & 3.91 & 4.80 & 5.10 & 0.91 \\ 
  \hline
  0.313E-5 & par & 0.71 & 1.04 & 1.14 & 1.17 & 1.42 & 1.51 & \bf 0.27 \\ 
  0.313E-5 & se & 0.18 & 0.20 & 0.22 & \bf 0.22 & 0.24 & 0.31 & 0.04 \\ 
  0.313E-5 & z & 3.67 & 4.67 & 5.33 & 5.24 & 5.81 & 6.39 & 0.86 \\ 
  \hline
  0.625E-5 & par & 1.06 & 1.20 & 1.31 & 1.33 & 1.46 & 1.63 & \bf 0.18 \\ 
  0.625E-5 & se & 0.13 & 0.15 & 0.16 & \bf 0.16 & 0.16 & 0.18 & 0.01 \\ 
  0.625E-5 & z & 7.52 & 8.00 & 8.52 & 8.50 & 9.02 & 9.60 & 0.71 \\ 
  \hline
  0.125E-4 & par & 1.11 & 1.20 & 1.29 & 1.28 & 1.34 & 1.46 & \bf 0.11 \\ 
  0.125E-4 & se & 0.10 & 0.10 & 0.11 & \bf 0.11 & 0.11 & 0.12 & 0.01 \\ 
  0.125E-4 & z & 10.11 & 11.46 & 12.09 & 11.75 & 12.22 & 12.95 & 0.92 \\ 
  \hline
  0.25E-4 & par & 1.18 & 1.22 & 1.25 & 1.28 & 1.31 & 1.40 & \bf 0.07 \\ 
  0.25E-4 & se & 0.07 & 0.07 & 0.07 & \bf 0.07 & 0.08 & 0.08 & 0.00 \\ 
  0.25E-4 & z & 15.70 & 16.62 & 17.30 & 17.13 & 17.47 & 18.83 & 0.88 \\ 
  \hline
  0.5E-4 & par & 1.21 & 1.28 & 1.31 & 1.31 & 1.35 & 1.39 & \bf 0.06 \\ 
  0.5E-4 & se & 0.05 & 0.05 & 0.05 & \bf 0.05 & 0.05 & 0.06 & 0.00 \\ 
  0.5E-4 & z & 23.33 & 24.66 & 25.13 & 25.01 & 25.60 & 25.90 & 0.80 \\ 
   \hline
\end{tabular}
\end{table}

\begin{table}
  \caption{Summary statistics of parameters (\emph{par}), standard errors (\emph{se}), and z-values (\emph{z}) for the \textbf{\emph{activity}} effect over 10 samples with $m=5$ and varying $p$ (as given in the first column).}
  \label{tab:summary_act_vary_prob}
\centering\begin{tabular}{l@{\quad}l@{\quad}r@{\quad}r@{\quad}r@{\quad}r@{\quad}r@{\quad}r@{\quad}r}
  \hline
$p$ &  & min & 1st qu & median & mean & 3rd qu & max & sd \\ 
  \hline
0.977E-7 & par & 0.49 & 5.69 & 10.61 & 10.63 & 12.11 & 30.82 & \bf 8.69 \\ 
  0.977E-7 & se & 0.56 & 103.33 & 3698.22 & \bf 2992.04 & 4814.90 & 7721.00 & 2744.08 \\ 
  0.977E-7 & z & 0.00 & 0.00 & 0.00 & 0.24 & 0.20 & 1.27 & 0.45 \\ 
  \hline
  0.195E-6 & par & 0.66 & 1.25 & 3.89 & 7.90 & 14.02 & 24.62 & \bf 8.95 \\ 
  0.195E-6 & se & 0.37 & 1.73 & 12.12 & \bf 1964.63 & 414.82 & 15140.75 & 4784.95 \\ 
  0.195E-6 & z & 0.00 & 0.07 & 0.30 & 0.67 & 1.36 & 2.00 & 0.81 \\ 
  \hline
  0.391E-6 & par & 1.11 & 1.24 & 1.61 & 2.25 & 2.60 & 6.64 & \bf 1.69 \\ 
  0.391E-6 & se & 0.24 & 0.44 & 0.69 & \bf 1.80 & 1.46 & 8.64 & 2.63 \\ 
  0.391E-6 & z & 0.37 & 1.74 & 2.40 & 2.37 & 2.84 & 4.54 & 1.11 \\ 
  \hline
  0.781E-6 & par & 0.97 & 1.25 & 1.61 & 1.84 & 2.09 & 3.53 & \bf 0.81 \\ 
  0.781E-6 & se & 0.23 & 0.31 & 0.41 & \bf 0.59 & 0.69 & 1.37 & 0.40 \\ 
  0.781E-6 & z & 2.35 & 2.70 & 3.74 & 3.59 & 4.29 & 4.78 & 0.92 \\ 
  \hline
  0.156E-5 & par & 1.12 & 1.47 & 1.67 & 1.78 & 1.97 & 2.61 & \bf 0.46 \\ 
  0.156E-5 & se & 0.22 & 0.25 & 0.30 & \bf 0.36 & 0.43 & 0.64 & 0.14 \\ 
  0.156E-5 & z & 3.69 & 4.59 & 5.54 & 5.25 & 5.73 & 6.72 & 0.96 \\ 
  \hline
  0.313E-5 & par & 1.28 & 1.55 & 1.70 & 1.73 & 1.83 & 2.35 & \bf 0.31 \\ 
  0.313E-5 & se & 0.16 & 0.20 & 0.24 & \bf 0.24 & 0.27 & 0.36 & 0.06 \\ 
  0.313E-5 & z & 6.44 & 6.76 & 7.25 & 7.30 & 7.86 & 8.08 & 0.62 \\ 
  \hline
  0.625E-5 & par & 1.32 & 1.43 & 1.52 & 1.56 & 1.62 & 1.94 & \bf 0.21 \\ 
  0.625E-5 & se & 0.11 & 0.12 & 0.13 & \bf 0.14 & 0.15 & 0.20 & 0.03 \\ 
  0.625E-5 & z & 9.66 & 11.03 & 11.47 & 11.41 & 12.14 & 12.56 & 0.92 \\ 
  \hline
  0.125E-4 & par & 1.39 & 1.42 & 1.51 & 1.51 & 1.60 & 1.67 & \bf 0.10 \\ 
  0.125E-4 & se & 0.08 & 0.09 & 0.09 & \bf 0.09 & 0.10 & 0.11 & 0.01 \\ 
  0.125E-4 & z & 14.23 & 15.88 & 16.13 & 16.09 & 16.54 & 17.39 & 0.91 \\ 
  \hline
  0.25E-4 & par & 1.41 & 1.46 & 1.52 & 1.51 & 1.55 & 1.58 & \bf 0.06 \\ 
  0.25E-4 & se & 0.06 & 0.06 & 0.06 & \bf 0.06 & 0.07 & 0.07 & 0.00 \\ 
  0.25E-4 & z & 22.93 & 23.08 & 23.40 & 23.46 & 23.52 & 25.10 & 0.62 \\ 
  \hline
  0.5E-4 & par & 1.45 & 1.46 & 1.50 & 1.52 & 1.59 & 1.63 & \bf 0.07 \\ 
  0.5E-4 & se & 0.04 & 0.04 & 0.04 & \bf 0.05 & 0.05 & 0.05 & 0.00 \\ 
  0.5E-4 & z & 31.86 & 32.90 & 33.73 & 33.56 & 34.07 & 34.77 & 0.91 \\ 
   \hline
\end{tabular}
\end{table}

\begin{table}
  \caption{Summary statistics of parameters (\emph{par}), standard errors (\emph{se}), and z-values (\emph{z}) for the \textbf{\emph{four-cycle}} effect over 10 samples with $m=5$ and varying $p$ (as given in the first column).}
  \label{tab:summary_4cy_vary_prob}
\centering\begin{tabular}{l@{\quad}l@{\quad}r@{\quad}r@{\quad}r@{\quad}r@{\quad}r@{\quad}r@{\quad}r}
  \hline
$p$ &  & min & 1st qu & median & mean & 3rd qu & max & sd \\ 
  \hline
0.977E-7 & par & -18.92 & -3.98 & 2.61 & 5.63 & 6.44 & 62.52 & \bf 21.94 \\ 
  0.977E-7 & se & 5.81 & 526.14 & 6998.87 & \bf 5963.65 & 7674.41 & 15535.57 & 5562.52 \\ 
  0.977E-7 & z & -0.01 & -0.00 & 0.00 & 0.09 & 0.04 & 0.52 & 0.19 \\ 
  \hline
  0.195E-6 & par & -28.76 & -0.04 & 3.74 & 19.84 & 53.04 & 93.26 & \bf 42.49 \\ 
  0.195E-6 & se & 0.96 & 24.15 & 75.29 & \bf 2309.34 & 3626.52 & 11803.40 & 4023.58 \\ 
  0.195E-6 & z & -0.36 & -0.00 & 0.02 & 0.08 & 0.27 & 0.54 & 0.27 \\ 
  \hline
  0.391E-6 & par & -8.50 & 0.86 & 2.26 & 8.42 & 4.33 & 70.53 & \bf 22.24 \\ 
  0.391E-6 & se & 0.76 & 1.92 & 4.49 & \bf 40.34 & 13.34 & 343.64 & 106.75 \\ 
  0.391E-6 & z & -1.50 & 0.19 & 0.40 & 0.32 & 0.84 & 1.39 & 0.83 \\ 
  \hline
  0.781E-6 & par & -1.23 & 0.16 & 0.24 & 2.99 & 4.48 & 15.70 & \bf 5.17 \\ 
  0.781E-6 & se & 0.57 & 0.76 & 1.08 & \bf 2.58 & 3.20 & 10.94 & 3.17 \\ 
  0.781E-6 & z & -1.83 & 0.19 & 0.26 & 0.39 & 1.37 & 2.00 & 1.21 \\ 
  \hline
  0.156E-5 & par & -0.39 & 0.56 & 1.73 & 2.43 & 2.93 & 8.22 & \bf 2.77 \\ 
  0.156E-5 & se & 0.42 & 1.08 & 1.23 & \bf 1.74 & 1.45 & 6.02 & 1.67 \\ 
  0.156E-5 & z & -0.85 & 0.46 & 1.27 & 1.08 & 1.97 & 2.08 & 1.01 \\ 
  \hline
  0.313E-5 & par & 0.21 & 0.52 & 1.43 & 1.98 & 3.21 & 4.98 & \bf 1.86 \\ 
  0.313E-5 & se & 0.57 & 0.70 & 0.80 & \bf 1.23 & 1.78 & 2.83 & 0.79 \\ 
  0.313E-5 & z & 0.37 & 0.80 & 1.27 & 1.38 & 1.91 & 2.54 & 0.74 \\ 
  \hline
  0.625E-5 & par & 0.25 & 0.69 & 1.08 & 1.19 & 1.77 & 2.04 & \bf 0.66 \\ 
  0.625E-5 & se & 0.25 & 0.36 & 0.47 & \bf 0.47 & 0.60 & 0.70 & 0.16 \\ 
  0.625E-5 & z & 0.90 & 1.95 & 2.51 & 2.37 & 2.88 & 3.37 & 0.81 \\ 
  \hline
  0.125E-4 & par & 0.33 & 0.74 & 0.79 & 0.89 & 1.12 & 1.36 & \bf 0.33 \\ 
  0.125E-4 & se & 0.25 & 0.27 & 0.34 & \bf 0.32 & 0.36 & 0.42 & 0.05 \\ 
  0.125E-4 & z & 0.90 & 2.45 & 2.85 & 2.77 & 2.98 & 3.93 & 0.86 \\ 
  \hline
  0.25E-4 & par & 0.56 & 0.66 & 0.98 & 0.87 & 1.02 & 1.11 & \bf 0.21 \\ 
  0.25E-4 & se & 0.16 & 0.20 & 0.21 & \bf 0.21 & 0.23 & 0.24 & 0.02 \\ 
  0.25E-4 & z & 2.75 & 3.76 & 4.18 & 4.07 & 4.55 & 5.01 & 0.74 \\ 
  \hline
  0.5E-4 & par & 0.50 & 0.67 & 0.70 & 0.76 & 0.93 & 1.04 & \bf 0.19 \\ 
  0.5E-4 & se & 0.12 & 0.13 & 0.13 & \bf 0.14 & 0.15 & 0.17 & 0.02 \\ 
  0.5E-4 & z & 3.99 & 5.11 & 5.42 & 5.53 & 5.86 & 7.48 & 0.98 \\ 
   \hline
\end{tabular}
\end{table}

\begin{table}
  \caption{Summary statistics of parameters (\emph{par}), standard errors (\emph{se}), and z-values (\emph{z}) for the \textbf{\emph{assortativity}} effect over 10 samples with $m=5$ and varying $p$ (as given in the first column).}
  \label{tab:summary_ass_vary_prob}
\centering\begin{tabular}{l@{\quad}l@{\quad}r@{\quad}r@{\quad}r@{\quad}r@{\quad}r@{\quad}r@{\quad}r}
  \hline
$p$ &  & min & 1st qu & median & mean & 3rd qu & max & sd \\ 
  \hline
0.977E-7 & par & -9.31 & -4.00 & -1.81 & -1.56 & 1.54 & 7.66 & \bf 4.91 \\ 
  0.977E-7 & se & 0.91 & 79.70 & 1588.41 & \bf 1614.18 & 2140.75 & 4405.15 & 1667.14 \\ 
  0.977E-7 & z & -0.72 & -0.00 & -0.00 & -0.01 & 0.00 & 0.61 & 0.32 \\ 
  \hline
  0.195E-6 & par & -13.44 & -1.76 & 0.50 & 2.36 & 11.23 & 13.31 & \bf 9.20 \\ 
  0.195E-6 & se & 0.48 & 7.36 & 39.79 & \bf 1545.68 & 425.53 & 11377.11 & 3609.04 \\ 
  0.195E-6 & z & -0.61 & -0.00 & 0.14 & 0.14 & 0.36 & 0.85 & 0.42 \\ 
  \hline
  0.391E-6 & par & -0.91 & -0.37 & 0.04 & 0.41 & 0.92 & 2.47 & \bf 1.13 \\ 
  0.391E-6 & se & 0.16 & 0.53 & 0.77 & \bf 2.14 & 2.48 & 10.56 & 3.17 \\ 
  0.391E-6 & z & -2.32 & -1.33 & 0.03 & -0.40 & 0.44 & 0.73 & 1.09 \\ 
  \hline
  0.781E-6 & par & -1.41 & -0.33 & 0.04 & 0.16 & 0.36 & 2.94 & \bf 1.22 \\ 
  0.781E-6 & se & 0.17 & 0.24 & 0.48 & \bf 0.61 & 0.68 & 1.84 & 0.53 \\ 
  0.781E-6 & z & -2.14 & -1.09 & 0.07 & -0.08 & 0.91 & 1.60 & 1.33 \\ 
  \hline
  0.156E-5 & par & -0.65 & -0.32 & -0.29 & -0.12 & -0.17 & 1.23 & \bf 0.52 \\ 
  0.156E-5 & se & 0.16 & 0.21 & 0.27 & \bf 0.33 & 0.33 & 0.86 & 0.21 \\ 
  0.156E-5 & z & -2.50 & -1.50 & -1.02 & -0.75 & -0.50 & 1.43 & 1.28 \\ 
  \hline
  0.313E-5 & par & -0.40 & -0.22 & -0.08 & -0.09 & 0.08 & 0.15 & \bf 0.19 \\ 
  0.313E-5 & se & 0.14 & 0.18 & 0.21 & \bf 0.22 & 0.27 & 0.35 & 0.07 \\ 
  0.313E-5 & z & -2.04 & -0.98 & -0.34 & -0.45 & 0.29 & 0.70 & 0.93 \\ 
  \hline
  0.625E-5 & par & -0.41 & -0.26 & -0.19 & -0.20 & -0.14 & 0.02 & \bf 0.12 \\ 
  0.625E-5 & se & 0.08 & 0.10 & 0.11 & \bf 0.11 & 0.13 & 0.16 & 0.02 \\ 
  0.625E-5 & z & -2.76 & -2.32 & -1.51 & -1.66 & -1.37 & 0.20 & 0.85 \\ 
  \hline
  0.125E-4 & par & -0.22 & -0.19 & -0.17 & -0.13 & -0.13 & 0.24 & \bf 0.13 \\ 
  0.125E-4 & se & 0.06 & 0.07 & 0.08 & \bf 0.08 & 0.08 & 0.13 & 0.02 \\ 
  0.125E-4 & z & -3.27 & -2.58 & -2.25 & -1.87 & -1.67 & 1.86 & 1.42 \\ 
  \hline
  0.25E-4 & par & -0.27 & -0.19 & -0.17 & -0.15 & -0.10 & 0.00 & \bf 0.08 \\ 
  0.25E-4 & se & 0.04 & 0.05 & 0.05 & \bf 0.05 & 0.06 & 0.06 & 0.01 \\ 
  0.25E-4 & z & -5.92 & -3.99 & -3.46 & -2.99 & -1.75 & 0.07 & 1.75 \\ 
  \hline
  0.5E-4 & par & -0.23 & -0.20 & -0.17 & -0.17 & -0.15 & -0.11 & \bf 0.04 \\ 
  0.5E-4 & se & 0.03 & 0.03 & 0.04 & \bf 0.04 & 0.04 & 0.04 & 0.00 \\ 
  0.5E-4 & z & -6.94 & -5.55 & -4.67 & -4.85 & -4.04 & -3.14 & 1.30 \\ 
   \hline
\end{tabular}
\end{table}


\begin{table}
  \caption{Summary statistics of parameters (\emph{par}), standard errors (\emph{se}), and z-values (\emph{z}) for the \textbf{\emph{repetition}} effect over 10 samples with $p=1.0E-5$ and varying $m$ (as given in the first column).}
  \label{tab:summary_rep_vary_ccs}
\centering\begin{tabular}{l@{\quad}l@{\quad}r@{\quad}r@{\quad}r@{\quad}r@{\quad}r@{\quad}r@{\quad}r}
  \hline
$m$ &  & min & 1st qu & median & mean & 3rd qu & max & sd \\ 
  \hline
1 & par & 65.46 & 76.66 & 86.95 & 84.80 & 91.64 & 101.02 & \bf 10.66 \\ 
  1 & se & 81.66 & 104.62 & 184.99 & \bf 183.31 & 258.16 & 311.05 & 89.50 \\ 
  1 & z & 0.24 & 0.36 & 0.50 & 0.57 & 0.78 & 0.93 & 0.25 \\ 
  \hline
  2 & par & 45.70 & 75.30 & 89.39 & 84.20 & 94.86 & 109.33 & \bf 19.18 \\ 
  2 & se & 32.85 & 76.29 & 147.94 & \bf 180.60 & 248.64 & 485.05 & 140.31 \\ 
  2 & z & 0.21 & 0.38 & 0.52 & 0.73 & 1.07 & 1.42 & 0.46 \\ 
  \hline
  4 & par & 33.99 & 68.32 & 80.18 & 74.08 & 87.87 & 94.42 & \bf 20.20 \\ 
  4 & se & 24.61 & 46.56 & 57.82 & \bf 79.83 & 73.77 & 291.75 & 77.35 \\ 
  4 & z & 0.32 & 0.95 & 1.03 & 1.34 & 1.60 & 3.26 & 0.81 \\ 
  \hline
  8 & par & 11.93 & 55.17 & 69.01 & 67.11 & 88.18 & 99.30 & \bf 26.24 \\ 
  8 & se & 2.37 & 35.12 & 48.82 & \bf 49.18 & 66.96 & 87.42 & 24.45 \\ 
  8 & z & 0.94 & 1.27 & 1.36 & 1.78 & 1.67 & 5.03 & 1.19 \\ 
  \hline
  16 & par & 10.45 & 37.27 & 72.21 & 63.85 & 84.24 & 105.82 & \bf 34.09 \\ 
  16 & se & 1.76 & 17.86 & 32.96 & \bf 33.71 & 54.28 & 60.01 & 22.34 \\ 
  16 & z & 1.36 & 1.66 & 1.89 & 2.60 & 3.06 & 5.93 & 1.47 \\ 
  \hline
  32 & par & 12.30 & 60.51 & 74.40 & 65.04 & 84.03 & 98.10 & \bf 29.47 \\ 
  32 & se & 1.41 & 18.53 & 24.73 & \bf 23.94 & 33.39 & 44.54 & 13.96 \\ 
  32 & z & 1.92 & 2.43 & 3.07 & 3.80 & 3.67 & 8.71 & 2.25 \\ 
  \hline
  64 & par & 19.85 & 25.22 & 44.10 & 47.24 & 67.03 & 79.21 & \bf 23.92 \\ 
  64 & se & 3.89 & 4.84 & 18.18 & \bf 22.68 & 31.18 & 64.24 & 20.99 \\ 
  64 & z & 1.23 & 2.08 & 2.65 & 3.31 & 4.68 & 6.71 & 1.77 \\ 
  \hline
  128 & par & 16.83 & 31.61 & 54.74 & 53.03 & 61.24 & 95.40 & \bf 25.90 \\ 
  128 & se & 2.59 & 6.27 & 14.40 & \bf 12.90 & 16.69 & 26.17 & 7.53 \\ 
  128 & z & 3.13 & 3.90 & 4.30 & 4.68 & 5.29 & 7.00 & 1.30 \\ 
   \hline
\end{tabular}
\end{table}

\begin{table}
  \caption{Summary statistics of parameters (\emph{par}), standard errors (\emph{se}), and z-values (\emph{z}) for the \textbf{\emph{popularity}} effect over 10 samples with $p=1.0E-5$ and varying $m$ (as given in the first column).}
  \label{tab:summary_pop_vary_ccs}
\centering\begin{tabular}{l@{\quad}l@{\quad}r@{\quad}r@{\quad}r@{\quad}r@{\quad}r@{\quad}r@{\quad}r}
  \hline
$m$ &  & min & 1st qu & median & mean & 3rd qu & max & sd \\ 
  \hline
1 & par & 1.03 & 1.33 & 1.38 & 1.35 & 1.45 & 1.55 & \bf 0.18 \\ 
  1 & se & 0.18 & 0.21 & 0.23 & \bf 0.23 & 0.26 & 0.30 & 0.04 \\ 
  1 & z & 4.87 & 5.47 & 5.89 & 5.80 & 6.25 & 6.38 & 0.53 \\ 
  \hline
  2 & par & 1.02 & 1.11 & 1.30 & 1.29 & 1.44 & 1.66 & \bf 0.21 \\ 
  2 & se & 0.13 & 0.15 & 0.16 & \bf 0.17 & 0.19 & 0.22 & 0.03 \\ 
  2 & z & 6.55 & 7.34 & 7.90 & 7.69 & 8.15 & 8.63 & 0.72 \\ 
  \hline
  4 & par & 1.10 & 1.19 & 1.28 & 1.31 & 1.45 & 1.53 & \bf 0.15 \\ 
  4 & se & 0.11 & 0.12 & 0.13 & \bf 0.12 & 0.13 & 0.14 & 0.01 \\ 
  4 & z & 8.92 & 9.83 & 10.37 & 10.50 & 11.35 & 11.71 & 0.94 \\ 
  \hline
  8 & par & 1.12 & 1.24 & 1.27 & 1.28 & 1.33 & 1.40 & \bf 0.08 \\ 
  8 & se & 0.10 & 0.10 & 0.10 & \bf 0.10 & 0.11 & 0.11 & 0.00 \\ 
  8 & z & 11.51 & 12.30 & 12.61 & 12.56 & 12.91 & 13.42 & 0.55 \\ 
  \hline
  16 & par & 1.13 & 1.17 & 1.24 & 1.26 & 1.35 & 1.42 & \bf 0.10 \\ 
  16 & se & 0.08 & 0.08 & 0.08 & \bf 0.08 & 0.09 & 0.09 & 0.00 \\ 
  16 & z & 13.43 & 14.32 & 14.87 & 15.13 & 16.01 & 17.06 & 1.14 \\ 
  \hline
  32 & par & 1.08 & 1.15 & 1.21 & 1.20 & 1.23 & 1.36 & \bf 0.08 \\ 
  32 & se & 0.06 & 0.07 & 0.07 & \bf 0.07 & 0.07 & 0.07 & 0.00 \\ 
  32 & z & 16.15 & 16.59 & 17.78 & 17.82 & 18.47 & 20.20 & 1.39 \\ 
  \hline
  64 & par & 1.15 & 1.22 & 1.26 & 1.25 & 1.29 & 1.32 & \bf 0.05 \\ 
  64 & se & 0.06 & 0.06 & 0.06 & \bf 0.06 & 0.06 & 0.06 & 0.00 \\ 
  64 & z & 18.19 & 20.90 & 21.58 & 21.19 & 21.98 & 22.17 & 1.22 \\ 
  \hline
  128 & par & 1.16 & 1.23 & 1.25 & 1.24 & 1.26 & 1.27 & \bf 0.03 \\ 
  128 & se & 0.05 & 0.05 & 0.05 & \bf 0.05 & 0.05 & 0.05 & 0.00 \\ 
  128 & z & 22.16 & 23.23 & 23.43 & 23.72 & 24.26 & 25.79 & 1.01 \\ 
   \hline
\end{tabular}
\end{table}

\begin{table}
  \caption{Summary statistics of parameters (\emph{par}), standard errors (\emph{se}), and z-values (\emph{z}) for the \textbf{\emph{activity}} effect over 10 samples with $p=1.0E-5$ and varying $m$ (as given in the first column).}
  \label{tab:summary_act_vary_ccs}
\centering\begin{tabular}{l@{\quad}l@{\quad}r@{\quad}r@{\quad}r@{\quad}r@{\quad}r@{\quad}r@{\quad}r}
  \hline
$m$ &  & min & 1st qu & median & mean & 3rd qu & max & sd \\ 
  \hline
1 & par & 1.21 & 1.59 & 1.68 & 1.75 & 1.84 & 2.64 & \bf 0.37 \\ 
  1 & se & 0.15 & 0.21 & 0.24 & \bf 0.25 & 0.26 & 0.46 & 0.08 \\ 
  1 & z & 5.71 & 6.84 & 7.32 & 7.27 & 7.77 & 8.28 & 0.78 \\ 
  \hline
  2 & par & 1.26 & 1.44 & 1.54 & 1.60 & 1.66 & 2.18 & \bf 0.27 \\ 
  2 & se & 0.11 & 0.13 & 0.15 & \bf 0.16 & 0.17 & 0.26 & 0.04 \\ 
  2 & z & 8.34 & 9.88 & 10.31 & 10.34 & 10.89 & 11.92 & 1.09 \\ 
  \hline
  4 & par & 1.35 & 1.43 & 1.54 & 1.55 & 1.64 & 1.84 & \bf 0.16 \\ 
  4 & se & 0.09 & 0.11 & 0.12 & \bf 0.12 & 0.12 & 0.14 & 0.01 \\ 
  4 & z & 12.40 & 12.93 & 13.25 & 13.33 & 13.67 & 14.58 & 0.61 \\ 
  \hline
  8 & par & 1.36 & 1.40 & 1.43 & 1.47 & 1.56 & 1.66 & \bf 0.11 \\ 
  8 & se & 0.07 & 0.08 & 0.08 & \bf 0.08 & 0.09 & 0.10 & 0.01 \\ 
  8 & z & 16.51 & 17.40 & 17.56 & 17.61 & 17.92 & 18.58 & 0.58 \\ 
  \hline
  16 & par & 1.32 & 1.39 & 1.44 & 1.45 & 1.49 & 1.62 & \bf 0.08 \\ 
  16 & se & 0.06 & 0.06 & 0.06 & \bf 0.06 & 0.07 & 0.07 & 0.00 \\ 
  16 & z & 21.74 & 21.98 & 22.36 & 22.45 & 22.80 & 23.36 & 0.56 \\ 
  \hline
  32 & par & 1.30 & 1.34 & 1.36 & 1.36 & 1.39 & 1.43 & \bf 0.04 \\ 
  32 & se & 0.04 & 0.05 & 0.05 & \bf 0.05 & 0.05 & 0.05 & 0.00 \\ 
  32 & z & 26.24 & 28.87 & 29.11 & 29.14 & 29.75 & 30.51 & 1.19 \\ 
  \hline
  64 & par & 1.32 & 1.35 & 1.36 & 1.36 & 1.38 & 1.42 & \bf 0.03 \\ 
  64 & se & 0.04 & 0.04 & 0.04 & \bf 0.04 & 0.04 & 0.04 & 0.00 \\ 
  64 & z & 34.21 & 34.63 & 35.03 & 35.19 & 35.40 & 37.14 & 0.84 \\ 
  \hline
  128 & par & 1.30 & 1.32 & 1.34 & 1.34 & 1.35 & 1.38 & \bf 0.02 \\ 
  128 & se & 0.03 & 0.03 & 0.03 & \bf 0.03 & 0.03 & 0.03 & 0.00 \\ 
  128 & z & 42.17 & 42.68 & 43.62 & 43.36 & 43.91 & 44.34 & 0.74 \\ 
   \hline
\end{tabular}
\end{table}

\begin{table}
  \caption{Summary statistics of parameters (\emph{par}), standard errors (\emph{se}), and z-values (\emph{z}) for the \textbf{\emph{four-cycle}} effect over 10 samples with $p=1.0E-5$ and varying $m$ (as given in the first column).}
  \label{tab:summary_4cy_vary_ccs}
\centering\begin{tabular}{l@{\quad}l@{\quad}r@{\quad}r@{\quad}r@{\quad}r@{\quad}r@{\quad}r@{\quad}r}
  \hline
$m$ &  & min & 1st qu & median & mean & 3rd qu & max & sd \\ 
  \hline
1 & par & 0.66 & 1.02 & 1.27 & 1.35 & 1.67 & 2.28 & \bf 0.50 \\ 
  1 & se & 0.59 & 0.77 & 0.84 & \bf 0.93 & 1.00 & 1.46 & 0.30 \\ 
  1 & z & 0.75 & 1.32 & 1.39 & 1.47 & 1.68 & 2.28 & 0.43 \\ 
  \hline
  2 & par & -0.57 & 0.13 & 0.49 & 0.61 & 0.95 & 2.31 & \bf 0.81 \\ 
  2 & se & 0.28 & 0.34 & 0.39 & \bf 0.45 & 0.51 & 0.88 & 0.18 \\ 
  2 & z & -2.03 & 0.36 & 1.14 & 1.06 & 1.99 & 3.32 & 1.52 \\ 
  \hline
  4 & par & 0.56 & 0.61 & 0.78 & 0.86 & 1.06 & 1.33 & \bf 0.28 \\ 
  4 & se & 0.31 & 0.36 & 0.36 & \bf 0.38 & 0.40 & 0.55 & 0.06 \\ 
  4 & z & 1.57 & 1.82 & 2.05 & 2.25 & 2.51 & 3.63 & 0.66 \\ 
  \hline
  8 & par & -0.20 & 0.64 & 0.83 & 0.78 & 0.98 & 1.43 & \bf 0.42 \\ 
  8 & se & 0.18 & 0.25 & 0.26 & \bf 0.26 & 0.29 & 0.34 & 0.05 \\ 
  8 & z & -1.07 & 2.60 & 3.34 & 2.85 & 3.56 & 4.25 & 1.49 \\ 
  \hline
  16 & par & 0.32 & 0.61 & 0.75 & 0.75 & 0.88 & 1.14 & \bf 0.24 \\ 
  16 & se & 0.14 & 0.17 & 0.19 & \bf 0.19 & 0.21 & 0.22 & 0.03 \\ 
  16 & z & 2.30 & 3.60 & 4.06 & 3.91 & 4.41 & 5.22 & 0.82 \\ 
  \hline
  32 & par & 0.46 & 0.54 & 0.62 & 0.67 & 0.80 & 0.94 & \bf 0.17 \\ 
  32 & se & 0.11 & 0.12 & 0.13 & \bf 0.13 & 0.14 & 0.15 & 0.01 \\ 
  32 & z & 3.79 & 4.56 & 5.05 & 5.10 & 5.38 & 6.85 & 0.86 \\ 
  \hline
  64 & par & 0.45 & 0.50 & 0.63 & 0.63 & 0.74 & 0.84 & \bf 0.14 \\ 
  64 & se & 0.08 & 0.09 & 0.10 & \bf 0.10 & 0.11 & 0.11 & 0.01 \\ 
  64 & z & 5.07 & 5.49 & 6.05 & 6.31 & 7.21 & 7.95 & 1.00 \\ 
  \hline
  128 & par & 0.45 & 0.49 & 0.53 & 0.52 & 0.55 & 0.56 & \bf 0.04 \\ 
  128 & se & 0.07 & 0.07 & 0.07 & \bf 0.07 & 0.07 & 0.08 & 0.00 \\ 
  128 & z & 6.53 & 6.85 & 7.12 & 7.17 & 7.55 & 7.78 & 0.47 \\ 
   \hline
\end{tabular}
\end{table}

\begin{table}
  \caption{Summary statistics of parameters (\emph{par}), standard errors (\emph{se}), and z-values (\emph{z}) for the \textbf{\emph{assortativity}} effect over 10 samples with $p=1.0E-5$ and varying $m$ (as given in the first column).}
  \label{tab:summary_ass_vary_ccs}
\centering\begin{tabular}{l@{\quad}l@{\quad}r@{\quad}r@{\quad}r@{\quad}r@{\quad}r@{\quad}r@{\quad}r}
  \hline
$m$ &  & min & 1st qu & median & mean & 3rd qu & max & sd \\ 
  \hline
1 & par & -0.67 & -0.51 & -0.34 & -0.33 & -0.14 & -0.02 & \bf 0.23 \\ 
  1 & se & 0.14 & 0.15 & 0.16 & \bf 0.19 & 0.25 & 0.29 & 0.06 \\ 
  1 & z & -3.56 & -2.72 & -1.74 & -1.80 & -0.80 & -0.07 & 1.25 \\ 
  \hline
  2 & par & -0.37 & -0.25 & -0.16 & -0.13 & -0.01 & 0.13 & \bf 0.17 \\ 
  2 & se & 0.09 & 0.10 & 0.11 & \bf 0.12 & 0.12 & 0.18 & 0.03 \\ 
  2 & z & -3.56 & -1.98 & -1.25 & -1.21 & -0.14 & 1.03 & 1.54 \\ 
  \hline
  4 & par & -0.30 & -0.24 & -0.19 & -0.17 & -0.13 & 0.10 & \bf 0.11 \\ 
  4 & se & 0.08 & 0.09 & 0.10 & \bf 0.10 & 0.11 & 0.12 & 0.01 \\ 
  4 & z & -3.42 & -2.36 & -2.26 & -1.83 & -1.28 & 0.90 & 1.25 \\ 
  \hline
  8 & par & -0.27 & -0.21 & -0.16 & -0.16 & -0.09 & -0.07 & \bf 0.07 \\ 
  8 & se & 0.05 & 0.06 & 0.07 & \bf 0.07 & 0.07 & 0.08 & 0.01 \\ 
  8 & z & -4.39 & -3.17 & -2.30 & -2.34 & -1.34 & -0.94 & 1.15 \\ 
  \hline
  16 & par & -0.19 & -0.15 & -0.12 & -0.11 & -0.09 & -0.01 & \bf 0.06 \\ 
  16 & se & 0.05 & 0.05 & 0.05 & \bf 0.05 & 0.05 & 0.06 & 0.00 \\ 
  16 & z & -4.04 & -2.83 & -2.33 & -2.30 & -1.67 & -0.20 & 1.20 \\ 
  \hline
  32 & par & -0.18 & -0.17 & -0.15 & -0.12 & -0.08 & -0.03 & \bf 0.06 \\ 
  32 & se & 0.03 & 0.03 & 0.04 & \bf 0.04 & 0.04 & 0.04 & 0.00 \\ 
  32 & z & -5.54 & -4.80 & -4.41 & -3.53 & -1.95 & -0.65 & 1.87 \\ 
  \hline
  64 & par & -0.21 & -0.15 & -0.13 & -0.13 & -0.11 & -0.07 & \bf 0.04 \\ 
  64 & se & 0.02 & 0.03 & 0.03 & \bf 0.03 & 0.03 & 0.03 & 0.00 \\ 
  64 & z & -7.35 & -5.11 & -4.54 & -4.60 & -3.69 & -2.36 & 1.43 \\ 
  \hline
  128 & par & -0.14 & -0.12 & -0.10 & -0.10 & -0.09 & -0.07 & \bf 0.02 \\ 
  128 & se & 0.02 & 0.02 & 0.02 & \bf 0.02 & 0.02 & 0.02 & 0.00 \\ 
  128 & z & -6.76 & -5.61 & -4.64 & -4.79 & -3.91 & -3.04 & 1.17 \\ 
   \hline
\end{tabular}
\end{table}


\begin{table}
  \caption{Summary statistics of parameters (\emph{par}), standard errors (\emph{se}), and z-values (\emph{z}) for the \textbf{\emph{repetition}} effect over 10 samples with varying $m$ (as given in the first column) and varying $p$ satisfying $(m+1)\cdot p=6\cdot 0.0001$.}
  \label{tab:summary_rep_co_vary}
\centering\begin{tabular}{l@{\quad}l@{\quad}r@{\quad}r@{\quad}r@{\quad}r@{\quad}r@{\quad}r@{\quad}r}
  \hline
$m$ &  & min & 1st qu & median & mean & 3rd qu & max & sd \\ 
  \hline
2 & par & 7.83 & 8.48 & 9.36 & 10.79 & 12.34 & 17.16 & \bf 3.33 \\ 
  2 & se & 0.48 & 0.56 & 0.61 & \bf 0.97 & 1.15 & 2.21 & 0.66 \\ 
  2 & z & 7.06 & 11.03 & 14.73 & 13.16 & 15.46 & 16.44 & 3.37 \\ 
  \hline
  4 & par & 10.61 & 12.11 & 12.72 & 15.73 & 14.32 & 41.14 & \bf 9.09 \\ 
  4 & se & 0.67 & 0.94 & 1.11 & \bf 1.86 & 1.52 & 8.14 & 2.23 \\ 
  4 & z & 5.06 & 10.20 & 11.76 & 11.06 & 12.53 & 15.81 & 3.02 \\ 
  \hline
  8 & par & 14.64 & 50.30 & 62.76 & 60.02 & 80.75 & 88.85 & \bf 26.06 \\ 
  8 & se & 1.63 & 10.53 & 14.27 & \bf 13.34 & 18.95 & 20.42 & 6.62 \\ 
  8 & z & 3.58 & 4.42 & 4.68 & 4.98 & 4.92 & 8.97 & 1.46 \\ 
  \hline
  16 & par & 12.42 & 30.66 & 57.97 & 53.13 & 71.88 & 91.63 & \bf 27.80 \\ 
  16 & se & 1.24 & 5.43 & 12.32 & \bf 10.87 & 13.31 & 23.00 & 7.22 \\ 
  16 & z & 3.98 & 4.61 & 5.34 & 6.01 & 6.10 & 10.12 & 2.25 \\ 
  \hline
  32 & par & 15.91 & 31.35 & 67.61 & 56.56 & 75.97 & 93.35 & \bf 26.47 \\ 
  32 & se & 2.46 & 9.96 & 16.87 & \bf 16.56 & 23.24 & 27.58 & 8.51 \\ 
  32 & z & 2.80 & 3.18 & 3.47 & 3.71 & 3.81 & 6.47 & 1.06 \\ 
  \hline
  64 & par & 13.17 & 32.18 & 61.98 & 55.70 & 77.97 & 88.49 & \bf 27.29 \\ 
  64 & se & 1.39 & 8.86 & 19.70 & \bf 19.68 & 26.28 & 49.80 & 14.27 \\ 
  64 & z & 1.78 & 2.82 & 3.05 & 3.82 & 3.91 & 9.44 & 2.17 \\ 
  \hline
  128 & par & 25.74 & 32.59 & 79.49 & 64.01 & 83.31 & 102.49 & \bf 29.76 \\ 
  128 & se & 5.26 & 13.71 & 23.93 & \bf 26.64 & 33.59 & 59.53 & 17.01 \\ 
  128 & z & 1.34 & 2.29 & 2.60 & 2.85 & 3.53 & 4.89 & 1.10 \\ 
  \hline
  256 & par & 7.13 & 37.28 & 64.14 & 54.71 & 77.15 & 80.65 & \bf 27.54 \\ 
  256 & se & 0.58 & 11.71 & 15.78 & \bf 21.27 & 17.94 & 71.37 & 20.76 \\ 
  256 & z & 0.74 & 2.68 & 3.65 & 4.31 & 5.13 & 12.32 & 3.20 \\ 
   \hline
\end{tabular}
\end{table}

\begin{table}
  \caption{Summary statistics of parameters (\emph{par}), standard errors (\emph{se}), and z-values (\emph{z}) for the \textbf{\emph{popularity}} effect over 10 samples with varying $m$ (as given in the first column) and varying $p$ satisfying $(m+1)\cdot p=6\cdot 0.0001$.}
  \label{tab:summary_pop_co_vary}
\centering\begin{tabular}{l@{\quad}l@{\quad}r@{\quad}r@{\quad}r@{\quad}r@{\quad}r@{\quad}r@{\quad}r}
  \hline
$m$ &  & min & 1st qu & median & mean & 3rd qu & max & sd \\ 
  \hline
2 & par & 1.23 & 1.28 & 1.30 & 1.30 & 1.34 & 1.38 & \bf 0.05 \\ 
  2 & se & 0.03 & 0.03 & 0.04 & \bf 0.04 & 0.04 & 0.04 & 0.00 \\ 
  2 & z & 35.52 & 36.37 & 36.40 & 36.44 & 36.77 & 37.30 & 0.57 \\ 
  \hline
  4 & par & 1.24 & 1.28 & 1.31 & 1.31 & 1.34 & 1.36 & \bf 0.04 \\ 
  4 & se & 0.04 & 0.04 & 0.04 & \bf 0.04 & 0.04 & 0.04 & 0.00 \\ 
  4 & z & 34.80 & 35.18 & 35.57 & 35.69 & 36.27 & 36.64 & 0.63 \\ 
  \hline
  8 & par & 1.25 & 1.26 & 1.28 & 1.29 & 1.32 & 1.36 & \bf 0.04 \\ 
  8 & se & 0.04 & 0.04 & 0.04 & \bf 0.04 & 0.04 & 0.04 & 0.00 \\ 
  8 & z & 31.82 & 32.25 & 32.89 & 32.85 & 33.27 & 33.90 & 0.72 \\ 
  \hline
  16 & par & 1.13 & 1.22 & 1.25 & 1.23 & 1.26 & 1.29 & \bf 0.05 \\ 
  16 & se & 0.04 & 0.04 & 0.04 & \bf 0.04 & 0.04 & 0.05 & 0.00 \\ 
  16 & z & 26.50 & 27.90 & 28.34 & 28.32 & 28.84 & 29.62 & 0.87 \\ 
  \hline
  32 & par & 1.14 & 1.21 & 1.22 & 1.22 & 1.26 & 1.29 & \bf 0.05 \\ 
  32 & se & 0.05 & 0.05 & 0.05 & \bf 0.05 & 0.05 & 0.05 & 0.00 \\ 
  32 & z & 21.87 & 22.75 & 23.39 & 23.70 & 25.07 & 25.50 & 1.39 \\ 
  \hline
  64 & par & 1.08 & 1.21 & 1.23 & 1.22 & 1.25 & 1.36 & \bf 0.08 \\ 
  64 & se & 0.06 & 0.06 & 0.06 & \bf 0.06 & 0.06 & 0.07 & 0.00 \\ 
  64 & z & 17.36 & 17.86 & 19.51 & 19.31 & 20.28 & 21.61 & 1.52 \\ 
  \hline
  128 & par & 1.03 & 1.22 & 1.25 & 1.24 & 1.27 & 1.37 & \bf 0.09 \\ 
  128 & se & 0.07 & 0.07 & 0.08 & \bf 0.08 & 0.08 & 0.08 & 0.00 \\ 
  128 & z & 12.76 & 16.00 & 16.31 & 16.07 & 16.82 & 17.32 & 1.28 \\ 
  \hline
  256 & par & 0.54 & 0.61 & 1.09 & 0.93 & 1.17 & 1.27 & \bf 0.30 \\ 
  256 & se & 0.08 & 0.09 & 0.11 & \bf 0.39 & 0.68 & 1.16 & 0.43 \\ 
  256 & z & 0.47 & 1.00 & 10.57 & 7.59 & 11.65 & 14.21 & 5.89 \\ 
   \hline
\end{tabular}
\end{table}

\begin{table}
  \caption{Summary statistics of parameters (\emph{par}), standard errors (\emph{se}), and z-values (\emph{z}) for the \textbf{\emph{activity}} effect over 10 samples with varying $m$ (as given in the first column) and varying $p$ satisfying $(m+1)\cdot p=6\cdot 0.0001$.}
  \label{tab:summary_act_co_vary}
\centering\begin{tabular}{l@{\quad}l@{\quad}r@{\quad}r@{\quad}r@{\quad}r@{\quad}r@{\quad}r@{\quad}r}
  \hline
$m$ &  & min & 1st qu & median & mean & 3rd qu & max & sd \\ 
  \hline
2 & par & 1.48 & 1.51 & 1.53 & 1.53 & 1.54 & 1.58 & \bf 0.03 \\ 
  2 & se & 0.03 & 0.03 & 0.03 & \bf 0.03 & 0.03 & 0.03 & 0.00 \\ 
  2 & z & 46.32 & 46.77 & 47.11 & 47.26 & 47.79 & 48.40 & 0.69 \\ 
  \hline
  4 & par & 1.48 & 1.49 & 1.51 & 1.52 & 1.53 & 1.59 & \bf 0.03 \\ 
  4 & se & 0.03 & 0.03 & 0.03 & \bf 0.03 & 0.03 & 0.03 & 0.00 \\ 
  4 & z & 46.85 & 47.25 & 47.65 & 47.66 & 47.86 & 48.62 & 0.57 \\ 
  \hline
  8 & par & 1.39 & 1.44 & 1.47 & 1.46 & 1.49 & 1.50 & \bf 0.03 \\ 
  8 & se & 0.03 & 0.03 & 0.03 & \bf 0.03 & 0.03 & 0.03 & 0.00 \\ 
  8 & z & 45.01 & 45.73 & 46.52 & 46.23 & 46.74 & 46.89 & 0.65 \\ 
  \hline
  16 & par & 1.38 & 1.39 & 1.42 & 1.42 & 1.44 & 1.48 & \bf 0.03 \\ 
  16 & se & 0.03 & 0.03 & 0.03 & \bf 0.03 & 0.03 & 0.03 & 0.00 \\ 
  16 & z & 41.82 & 42.90 & 43.23 & 43.31 & 43.64 & 45.70 & 1.07 \\ 
  \hline
  32 & par & 1.36 & 1.40 & 1.42 & 1.41 & 1.42 & 1.46 & \bf 0.03 \\ 
  32 & se & 0.04 & 0.04 & 0.04 & \bf 0.04 & 0.04 & 0.04 & 0.00 \\ 
  32 & z & 36.55 & 37.97 & 38.64 & 38.31 & 38.91 & 39.63 & 1.04 \\ 
  \hline
  64 & par & 1.30 & 1.34 & 1.37 & 1.37 & 1.39 & 1.44 & \bf 0.04 \\ 
  64 & se & 0.04 & 0.04 & 0.04 & \bf 0.04 & 0.04 & 0.04 & 0.00 \\ 
  64 & z & 32.46 & 32.73 & 33.45 & 33.68 & 34.21 & 35.65 & 1.16 \\ 
  \hline
  128 & par & 1.27 & 1.32 & 1.33 & 1.34 & 1.37 & 1.43 & \bf 0.05 \\ 
  128 & se & 0.04 & 0.04 & 0.05 & \bf 0.05 & 0.05 & 0.05 & 0.00 \\ 
  128 & z & 28.02 & 28.17 & 29.28 & 29.31 & 30.41 & 30.65 & 1.13 \\ 
  \hline
  256 & par & 1.26 & 1.28 & 1.30 & 2.86 & 5.14 & 5.36 & \bf 2.03 \\ 
  256 & se & 0.05 & 0.05 & 0.05 & \bf 0.44 & 0.93 & 1.17 & 0.51 \\ 
  256 & z & 4.34 & 5.62 & 24.16 & 17.24 & 25.27 & 27.72 & 10.44 \\ 
   \hline
\end{tabular}
\end{table}

\begin{table}
  \caption{Summary statistics of parameters (\emph{par}), standard errors (\emph{se}), and z-values (\emph{z}) for the \textbf{\emph{four-cycle}} effect over 10 samples with varying $m$ (as given in the first column) and varying $p$ satisfying $(m+1)\cdot p=6\cdot 0.0001$.}
  \label{tab:summary_4cy_co_vary}
\centering\begin{tabular}{l@{\quad}l@{\quad}r@{\quad}r@{\quad}r@{\quad}r@{\quad}r@{\quad}r@{\quad}r}
  \hline
$m$ &  & min & 1st qu & median & mean & 3rd qu & max & sd \\ 
  \hline
2 & par & 0.45 & 0.61 & 0.70 & 0.72 & 0.85 & 1.01 & \bf 0.17 \\ 
  2 & se & 0.09 & 0.09 & 0.10 & \bf 0.10 & 0.11 & 0.11 & 0.01 \\ 
  2 & z & 5.20 & 6.20 & 6.94 & 7.06 & 7.90 & 8.91 & 1.20 \\ 
  \hline
  4 & par & 0.58 & 0.72 & 0.79 & 0.79 & 0.86 & 0.95 & \bf 0.12 \\ 
  4 & se & 0.10 & 0.10 & 0.10 & \bf 0.10 & 0.10 & 0.11 & 0.00 \\ 
  4 & z & 6.04 & 7.29 & 8.15 & 7.92 & 8.66 & 9.25 & 1.04 \\ 
  \hline
  8 & par & 0.60 & 0.71 & 0.77 & 0.75 & 0.80 & 0.91 & \bf 0.09 \\ 
  8 & se & 0.09 & 0.10 & 0.10 & \bf 0.10 & 0.10 & 0.11 & 0.01 \\ 
  8 & z & 6.58 & 7.40 & 7.94 & 7.68 & 8.08 & 8.39 & 0.60 \\ 
  \hline
  16 & par & 0.53 & 0.63 & 0.67 & 0.67 & 0.70 & 0.91 & \bf 0.10 \\ 
  16 & se & 0.09 & 0.09 & 0.10 & \bf 0.10 & 0.10 & 0.11 & 0.01 \\ 
  16 & z & 5.85 & 6.68 & 6.93 & 6.96 & 7.13 & 8.43 & 0.71 \\ 
  \hline
  32 & par & 0.47 & 0.59 & 0.63 & 0.63 & 0.66 & 0.80 & \bf 0.09 \\ 
  32 & se & 0.09 & 0.10 & 0.10 & \bf 0.10 & 0.10 & 0.12 & 0.01 \\ 
  32 & z & 5.28 & 5.64 & 6.53 & 6.34 & 6.71 & 7.42 & 0.77 \\ 
  \hline
  64 & par & 0.42 & 0.50 & 0.60 & 0.58 & 0.67 & 0.72 & \bf 0.10 \\ 
  64 & se & 0.08 & 0.10 & 0.10 & \bf 0.10 & 0.11 & 0.11 & 0.01 \\ 
  64 & z & 4.88 & 5.14 & 5.77 & 5.76 & 6.29 & 6.77 & 0.68 \\ 
  \hline
  128 & par & 0.20 & 0.44 & 0.57 & 0.53 & 0.63 & 0.75 & \bf 0.16 \\ 
  128 & se & 0.10 & 0.10 & 0.11 & \bf 0.11 & 0.12 & 0.12 & 0.01 \\ 
  128 & z & 2.09 & 4.24 & 5.08 & 4.78 & 5.53 & 6.26 & 1.18 \\ 
  \hline
  256 & par & 0.40 & 0.53 & 0.61 & 12.26 & 29.21 & 31.43 & \bf 15.16 \\ 
  256 & se & 0.11 & 0.12 & 0.13 & \bf 2.49 & 5.29 & 7.41 & 3.12 \\ 
  256 & z & 3.41 & 4.24 & 4.77 & 4.69 & 5.14 & 5.93 & 0.72 \\ 
   \hline
\end{tabular}
\end{table}

\begin{table}
  \caption{Summary statistics of parameters (\emph{par}), standard errors (\emph{se}), and z-values (\emph{z}) for the \textbf{\emph{assortativity}} effect over 10 samples with varying $m$ (as given in the first column) and varying $p$ satisfying $(m+1)\cdot p=6\cdot 0.0001$.}
  \label{tab:summary_ass_co_vary}
\centering\begin{tabular}{l@{\quad}l@{\quad}r@{\quad}r@{\quad}r@{\quad}r@{\quad}r@{\quad}r@{\quad}r}
  \hline
$m$ &  & min & 1st qu & median & mean & 3rd qu & max & sd \\ 
  \hline
2 & par & -0.23 & -0.20 & -0.19 & -0.18 & -0.17 & -0.13 & \bf 0.03 \\ 
  2 & se & 0.02 & 0.03 & 0.03 & \bf 0.03 & 0.03 & 0.03 & 0.00 \\ 
  2 & z & -8.96 & -7.55 & -7.16 & -6.91 & -6.21 & -4.57 & 1.38 \\ 
  \hline
  4 & par & -0.19 & -0.18 & -0.17 & -0.16 & -0.15 & -0.12 & \bf 0.03 \\ 
  4 & se & 0.02 & 0.03 & 0.03 & \bf 0.03 & 0.03 & 0.03 & 0.00 \\ 
  4 & z & -7.61 & -7.04 & -6.49 & -6.33 & -5.52 & -4.40 & 1.07 \\ 
  \hline
  8 & par & -0.21 & -0.16 & -0.13 & -0.14 & -0.11 & -0.11 & \bf 0.04 \\ 
  8 & se & 0.02 & 0.02 & 0.03 & \bf 0.03 & 0.03 & 0.03 & 0.00 \\ 
  8 & z & -8.77 & -6.52 & -5.21 & -5.56 & -4.10 & -3.92 & 1.65 \\ 
  \hline
  16 & par & -0.14 & -0.13 & -0.12 & -0.11 & -0.10 & -0.07 & \bf 0.02 \\ 
  16 & se & 0.02 & 0.03 & 0.03 & \bf 0.03 & 0.03 & 0.03 & 0.00 \\ 
  16 & z & -5.33 & -5.00 & -4.45 & -4.33 & -4.10 & -2.73 & 0.89 \\ 
  \hline
  32 & par & -0.20 & -0.13 & -0.10 & -0.11 & -0.08 & -0.06 & \bf 0.04 \\ 
  32 & se & 0.03 & 0.03 & 0.03 & \bf 0.03 & 0.03 & 0.03 & 0.00 \\ 
  32 & z & -7.30 & -4.63 & -3.79 & -3.92 & -2.68 & -1.69 & 1.65 \\ 
  \hline
  64 & par & -0.16 & -0.14 & -0.10 & -0.10 & -0.07 & -0.04 & \bf 0.04 \\ 
  64 & se & 0.03 & 0.03 & 0.03 & \bf 0.03 & 0.03 & 0.03 & 0.00 \\ 
  64 & z & -5.71 & -4.58 & -3.24 & -3.54 & -2.63 & -1.25 & 1.51 \\ 
  \hline
  128 & par & -0.17 & -0.14 & -0.08 & -0.10 & -0.07 & -0.05 & \bf 0.05 \\ 
  128 & se & 0.03 & 0.03 & 0.03 & \bf 0.03 & 0.03 & 0.04 & 0.00 \\ 
  128 & z & -5.36 & -4.29 & -2.68 & -3.16 & -2.00 & -1.33 & 1.45 \\ 
  \hline
  256 & par & -2.31 & -1.71 & -0.14 & -0.82 & -0.10 & -0.05 & \bf 0.94 \\ 
  256 & se & 0.03 & 0.03 & 0.04 & \bf 0.19 & 0.34 & 0.54 & 0.20 \\ 
  256 & z & -5.75 & -4.73 & -3.40 & -3.66 & -2.74 & -1.52 & 1.43 \\ 
   \hline
\end{tabular}
\end{table}

\end{document}